\PassOptionsToPackage{dvipsnames}{xcolor}
\documentclass[acmsmall]{acmart}

\setcopyright{cc}
\setcctype{by-nc-nd}
\acmDOI{10.1145/3839518}
\acmYear{2026}
\acmJournal{PACMPL}
\acmVolume{10}
\acmNumber{OOPSLA2}
\acmArticle{386}
\acmMonth{10}
\acmSubmissionID{oopslab26main-p1136-p}
\received{2026-03-17}
\received[accepted]{2026-08-06}

\usepackage{booktabs}   %
\usepackage{subcaption} %
\usepackage[dvipsnames]{xcolor}
\usepackage{bcprules}
\usepackage{collcell}
\usepackage{listings}
\usepackage[scaled=0.8]{beramono}
\usepackage{lipsum}
\usepackage{mathpartir}

\usepackage{mathtools}
\usepackage{mdframed}
\usepackage{amsmath}
\usepackage{varioref}
\usepackage{cleveref}
\usepackage{scalerel}
\usepackage{xspace}
\usepackage{calc}
\usepackage{array,varwidth}
\usepackage{makecell}
\usepackage{diagbox}
\usepackage{float}
\usepackage{xfrac}
\usepackage{enumitem}
\usepackage[skip=2.5pt]{caption}
\usepackage{slashbox}
\usepackage{xifthen}
\usepackage{tikz}
\usepackage{tikz-cd}
\usetikzlibrary{arrows}
\usepackage{cancel}
\usepackage{bm}
\usepackage{multicol}
\usepackage[export]{adjustbox}
\usepackage{longfbox}
\usepackage{stmaryrd}

\usetikzlibrary{shadows}
\usetikzlibrary{shadows.blur}
\usetikzlibrary{decorations.pathreplacing} %
\tikzcdset{arrow style=tikz,
           diagrams={>=stealth}
         }    %
\tikzcdset{crossing over clearance=.75ex}
\tikzset{hardd/.style={teal,thick}} %
\tikzset{softd/.style={teal, dashed,thick}} %
\tikzset{datad/.style={black,densely dotted}} %
\tikzset{alld/.style={black,thick}} %
\tikzset{expnode/.style={
    asymmetrical rectangle,rounded corners,draw,fill=white,inner sep=2.5pt,
    font=\footnotesize, %
  }} %
\tikzset{blknode/.style={
    asymmetrical rectangle,sharp corners,draw,violet,fill=violet!10,
    font=\footnotesize %
  }} %
\tikzset{dummynode/.style={fill=none,draw=none,asymmetrical rectangle} %
} %
\tikzset{nmute/.style={opacity=.4,blur shadow={shadow opacity=.4}}} %
\tikzset{expn/.style={
    circle,fill=white,draw,minimum size=5pt,inner sep=0pt,outer sep=0pt,
    font=\footnotesize, %
    label={[font=\footnotesize]left:#1}
  }} %
\tikzset{expn/.default={}}
\tikzset{mexpn/.style={opacity=.4,
    circle,fill=white,draw,minimum size=5pt,inner sep=0pt,outer sep=0pt,
    font=\footnotesize, %
    label={[font=\footnotesize,opacity=.4]left:#1}
  }} %
\tikzset{mexpn/.default={}}
\tikzset{rmexpn/.style={opacity=.4,
    circle,fill=white,draw,minimum size=5pt,inner sep=0pt,outer sep=0pt,
    font=\footnotesize, %
    label={[font=\footnotesize,opacity=.4]right:#1}
  }} %
\tikzset{rmexpn/.default={}}
\tikzset{mexpndummy/.style={opacity=.4,
    circle,fill=white,minimum size=5pt,inner sep=0pt,outer sep=0pt,
    label={[font=\footnotesize,opacity=.4]left:#1}
  }} %
\tikzset{mexpndummy/.default={}}

\makeatletter %
\def\arcr{\@arraycr}
\makeatother

\newcommand{\showDOI}[1]{\unskip}

\newtheoremstyle{mystyle}
  {1pt}    %
  {1pt}    %
  {\itshape}  %
  {\parindent}    %
  {} %
  {.}      %
  {5pt}    %
  {\thmname{#1}\thmnumber{ #2}\thmnote{ (#3)}} %

\theoremstyle{mystyle}
\newtheorem{re-theorem}{Theorem}[section]
\newtheorem{re-lemma}[re-theorem]{Lemma}
\newtheorem{re-corollary}[re-theorem]{Corollary}
\newtheorem{re-definition}[re-theorem]{Definition}

\newcommand{\figref}[1]{Fig.~\ref{#1}}
\newcommand{\secref}[1]{Section~\ref{#1}} %

\newcommand{\lemref}[1]{Lemma~\ref{#1}}

\newcommand{\defref}[1]{Def.~\ref{#1}}

\newcommand{\fr}{\textbf{\textit{fr}}}

\newcommand{\fv}{\textit{FV}}

\newcommand{\Specsharp}{%
	{\settoheight{\dimen0}{C}Spec\kern-.05em \resizebox{!}{\dimen0}{\raisebox{\depth}{\#}}}}
\newcommand{\Csharp}{%
	{\settoheight{\dimen0}{C}C\kern-.05em \resizebox{!}{\dimen0}{\raisebox{\depth}{\#}}}}

\newcommand{\fun}[1]{\operatorname{#1}}
\newcommand{\DOM}{\fun{dom}}

\definecolor{blue-violet}{rgb}{0.54, 0.17, 0.89}
\definecolor{depmap}{HTML}{00007B}
\definecolor{dark-cyan}{HTML}{135579}
\definecolor{magenta}{HTML}{a8264f}

\lstdefinelanguage{Neutral}%
{morekeywords={abstract,%
  case,catch,char,class,%
  def,else,extends,final,finally,for,%
  if,import,implicit,%
  match,module,%
  new,null,%
  object,override,%
  package,private,protected,public,%
  for,public,return,super,%
  this,trait,try,type,%
  val,var,%
  with,while,%
  yield,%
  let,end,%
	in,fun,alloc,inc%
  },%
  mathescape=true,%
  sensitive,%
  keywordstyle={\color{black}\bf\ttfamily},%
  commentstyle=\color{OliveGreen},%
  escapebegin=\color{OliveGreen},
  morecomment=[l]//,%
  morecomment=[s]{/*}{*/},%
  morecomment=[s][\color{darkgray}]{@}{\ },%
  morestring=[b]",%
  morestring=[b]',%
  showstringspaces=false%
}[keywords,comments,strings]%

\lstdefinelanguage{OOPSLA21}%
{morekeywords={abstract,%
  case,catch,char,class,%
  def,else,extends,final,finally,for,%
  if,import,implicit,%
  match,module,%
  new,null,%
  object,override,%
  package,private,protected,public,%
  for,public,return,super,%
  this,throw,trait,try,type,%
  val,var,%
  with,while,%
  yield,%
  let,end,%
	in,fun,alloc,inc%
  },%
  mathescape=true,%
  sensitive,%
  keywordstyle={\color{magenta}\bf\ttfamily},%
  commentstyle=\color{magenta},%
  escapebegin=\color{magenta},
  morecomment=[l]//,%
  morecomment=[s]{/*}{*/},%
  morecomment=[s][\color{magenta}]{@}{\ },%
  morestring=[b]",%
  morestring=[b]',%
  showstringspaces=false%
}[keywords,comments,strings]%

\lstdefinelanguage{PolyRT}%
{morekeywords=[1]{abstract,fn,%
  case,catch,char,class,%
  def,else,extends,final,finally,for,%
  if,import,implicit,%
  match,module,%
  new,null,%
  object,override,%
  package,private,protected,public,%
  for,public,return,super,%
  this,throw,trait,try,type,%
  val,var,%
  with,while,%
  yield,%
  let,end,%
	in,fun,%
  move,%
  },%
  keywords=[2]{%
    @eff,@cap,@fresh%
  },%
  mathescape=true,%
  sensitive,%
  keywordstyle=[1]{\color{dark-cyan}\bf\ttfamily},%
  keywordstyle=[2]{\color{dark-cyan}\ttfamily},%
  commentstyle=\color{dark-cyan},%
  escapebegin=\color{dark-cyan},%
  morecomment=[l]//,%
  morecomment=[s]{/*}{*/},%
  morestring=[b]",%
  morestring=[b]',%
  showstringspaces=false%
}[keywords,comments,strings]%

\newcommand{\Type}[1]{\ensuremath{\mathsf{#1}}}

\newcommand{\TRef}{\Type{Ref}}

\newcommand{\tref}{\mathsf{ref}}

\newcommand{\tlet}{\mathsf{let}}

\newcommand{\ty}[2][]{\ensuremath{\ifthenelse{\isempty{#1}}{#2}{#2^{\,#1}}}}

\newcommand{\ts}[1][]{\ensuremath{\ifthenelse{\isempty{#1}}{\,\vdash\,}{\,\vdash^{\,#1}\,}}}

\newcommand{\flt}{\ensuremath{\varphi}}

\newcommand{\cx}[2][]{\ensuremath{\ifthenelse{\isempty{#1}}{#2}{#2^{\,#1}}}}

\providecommand{\G}{G} %
\renewcommand{\G}[1][]{\cx[#1]{\Gamma}}

\newcommand{\qbot}{\ensuremath{\varnothing}}
\newcommand{\qfresh}{\ensuremath{\vardiamondsuit}}

\newsavebox{\SMALLSTAR}
\savebox{\SMALLSTAR}{\(\raisebox{.25ex}{\(\qfresh\)}\)}

\newsavebox{\OVRLP}
\savebox{\OVRLP}{$\raisebox{.37ex}[0pt][0pt]{$\mathrlap{\hspace{.415ex}\scaleobj{.5}{\vardiamondsuit}}$}\cap$}

\newcommand{\qsat}[1]{\ensuremath{#1\mathord{*}}}

\newfboxstyle{mybox}{padding=2pt,margin=2pt,baseline-skip=false}
\newcommand{\BOX}[1]{\lfbox[mybox]{\strut \footnotesize \ensuremath{#1}}}

\newcommand{\FV}{\ensuremath{\operatorname{fv}}}

\newcommand{\hole}[1]{\ensuremath{[\,#1\,]}}
\newcommand{\CX}[3][black]{\ensuremath{{\color{#1}#2\ifthenelse{\isempty{#3}}{}{\hole{{\color{black}#3}}}}}}

\newcommand{\ie}{{\em i.e.}\xspace}

\makeatletter
\newcommand{\artifacturl}[1]{
  \if@ACM@anonymous
    Link to repository removed for double-blind review.
  \else
    \url{#1}
  \fi
}
\makeatother

\newcommand{\bfparagraph}[1]{\paragraph{\textbf{#1}}}

\colorlet{eff}{dark-cyan}
\newcommand{\FX}[1]{\ensuremath{{\color{eff}#1}}}
\newcommand{\EPS}[1][]{\ifthenelse{\isempty{#1}}{\FX{\bm{\varepsilon}}}{\FX{\bm{\varepsilon_{#1}}}}}
\newcommand{\EPSS}[1][]{\ifthenelse{\isempty{#1}}{\FX{\qsat{\bm{\varepsilon}}}}{\FX{\qsat{\bm{\varepsilon_{#1}}}}}}
\newcommand{\EPSPR}[1][]{\ifthenelse{\isempty{#1}}{\FX{\bm{\varepsilon'}}}{\FX{\bm{\varepsilon'_{#1}}}}}
\newcommand{\EPSSPR}[1][]{\ifthenelse{\isempty{#1}}{\FX{\qsat{\bm{\varepsilon'}}}}{\FX{\qsat{\bm{\varepsilon'_{#1}}}}}}

\colorlet{meff}{Gray}
\newcommand{\M}{\mathnormal{m}}
\newcommand{\MFX}[1]{\ensuremath{{\color{meff}#1}}}
\newcommand{\MV}[1][]{\ifthenelse{\isempty{#1}}{\MFX{\bm{\M}}}{\MFX{\bm{\M_{#1}}}}}

\newcommand{\ccup}[1]{\ensuremath{\cup_{{#1}}}}

\colorlet{mute}{teal}

\newcommand{\ttrue}{\texttt{true}}
\newcommand{\tfalse}{\texttt{false}}
\newcommand{\tvar}[1]{#1}
\renewcommand{\tref}[1]{\texttt{ref}\ #1}
\newcommand{\tget}[1]{\texttt{!}#1}
\newcommand{\tput}[2]{#1\ \texttt{:=}\ #2}
\newcommand{\tapp}[2]{#1\ #2}
\newcommand{\tabs}[1]{\lambda x.\ #1}
\newcommand{\tnil}{\texttt{nil}}
\newcommand{\tcons}[2]{#1 \!::\! #2}
\newcommand{\tfold}[3]{\texttt{fold} \ #1 \ #2 \ #3 }

\renewcommand{\tlet}[3]{\texttt{let}\ #1 = #2\ \texttt{in}\ #3}

\newcommand{\nand}{\ensuremath{\barwedge}}
\newcommand{\tseqb}[2]{\ensuremath{#1 \nand #2}}

\newcommand{\qempty}{\emptyset}

\newcommand{\bfalse}{\bot}
\newcommand{\btrue}{\top}

\newcommand{\extends}{\ensuremath{\ensuremath{;}}}
\newcommand{\config}[3]{\ensuremath{{#1},\ {#2},\ {#3}}}
\newcommand{\pconfig}[2]{\ensuremath{{#1},\ {#2}}}
\newcommand{\cll}[2]{\ensuremath{\langle {#1}, \ {#2} \rangle }}
\newcommand{\eval}{\Downarrow}

\newcommand{\TE}{T}
\newcommand{\TEP}{T'}
\newcommand{\GE}{\Gamma}
\newcommand{\GEP}{\Gamma'}

\newcommand{\bfalseE}{\textcolor{red}{\bot}}
\newcommand{\btrueE}{\textcolor{red}{\top}}
\newcommand{\EF}[1]{\textcolor{red}{#1}}
\newcommand{\EFC}{\ensuremath{\textcolor{red}{\rhd}}}

\newcommand{\hasTypeE}[4]{#1 \vdash_{\rm{E}} #2 : #3\ \EF{#4}}
\newcommand{\indexrTypeE}[3]{#1(#2) = #3}

\newcommand{\TFunE}[3]{#1 \stackrel{\textcolor{red}{#3}}{\to} #2}
\newcommand{\TBoolE}{\texttt{Bool}}
\newcommand{\TRefE}{\texttt{Ref}}
\newcommand{\subtypeTE}[2]{#1\ <: #2}
\newcommand{\subtypeEE}[2]{\textcolor{red}{#1\ <: #2}}

\newcommand{\semTypeE}[5]{#1 \models_{\rm{E}} #2 \equivlog #3 : #4\ \textcolor{red}{#5}}
\newcommand{\ETE}[2]{\ensuremath{\mathcal{E}\synbracket{#1\ \textcolor{red}{#2}}}}
\newcommand{\ctxEquivE}[5]{#1 \models_{\rm{E}} #2 \equiva #3 : #4\ \textcolor{red}{#5}}

\newcommand{\TA}{T}
\newcommand{\TAP}{T'}
\newcommand{\GA}{\Gamma}
\newcommand{\GAP}{\Gamma'}

\newcommand{\bfalseA}{\textcolor{blue}{\bot}}
\newcommand{\btrueA}{\textcolor{blue}{\top}}
\newcommand{\AB}[1]{\textcolor{blue}{#1}}

\newcommand{\hasTypeA}[4]{#1 \vdash_{\rm{A}} #2 : #3\ \textcolor{blue}{#4}}
\newcommand{\indexrTypeA}[4]{#1(#2) = #3\ #4}
\renewcommand{\fv}[1]{\mathrm{fv}(#1)}
\newcommand{\envCapA}[3]{\textcolor{blue}{#1(#2) <: #3}}

\newcommand{\envCapTrueA}[2]{#1(#2) \ne \bfalse}
\newcommand{\hasTypePhiA}[5]{#1\ \textcolor{blue}{#2} \vdash_{\rm{A}} #3 : #4\ \textcolor{blue}{#5}}

\newcommand{\TFunA}[4]{#1\ \textcolor{blue}{#2} \to #3\ \textcolor{blue}{#4}}
\newcommand{\TBoolA}{\texttt{Bool}}
\newcommand{\TRefA}{\texttt{Ref}}

\newcommand{\subtypeTA}[2]{#1 <: #2}
\newcommand{\subtypeAA}[2]{\textcolor{blue}{#1 <: #2}}

\newcommand{\semTypeA}[5]{#1 \models_{\rm{A}} #2 \equivlog #3 : #4\ \textcolor{blue}{#5}}
\newcommand{\ETA}[2]{\ensuremath{\mathcal{E}\synbracket{#1\ \textcolor{blue}{#2}}}}
\newcommand{\ctxEquivA}[5]{#1 \models_{\rm{A}} #2 \equiva #3 : #4\ \textcolor{blue}{#5}}

\newcommand{\hasTypeAE}[5]{#1 \vdash #2 : #3\ \ #4\ #5}
\newcommand{\hasTypeAEC}[5]{#1 \vdash #2 : #3\ \ \textcolor{blue}{#4}\ \textcolor{red}{#5}}
\newcommand{\TBoolAE}{\texttt{Bool}}
\newcommand{\TRefAE}{\texttt{Ref}}
\newcommand{\TFunAE}[5]{#1\ \textcolor{blue}{#2} \stackrel{\textcolor{red}{#5}}{\to} #3\ \textcolor{blue}{#4}}
\newcommand{\TListAE}[1]{\texttt{List}\ #1}

\newcommand{\av}[2]{\langle #1, #2 \rangle}
\newcommand{\avc}[2]{\textcolor{blue}{\langle #1,\ #2 \rangle}}
\newcommand{\avbot}{\textcolor{blue}{\Bot}}
\newcommand{\avtop}{\textcolor{blue}{\Top}}

\newcommand{\trTA}[1]{\synbracket{#1}^a_{\rm{ae}}}
\newcommand{\trTE}[1]{\synbracket{#1}^e_{\rm{ae}}}

\newcommand{\DEF}{\stackrel{{\rm def}}{=}}

\newcommand{\st}[5]{\ensuremath{\scalebox{0.9}{\boxed{\!(#1, #2): #3\!}}}^{~ \scriptstyle #4}_{~ \scriptstyle #5}}

\newcommand{\VE}{\ensuremath{\hat{H}}}
\newcommand{\R}{\ensuremath{\hat{R}}}
\newcommand{\W}{\texttt{W}}
\newcommand{\stcp}[4]{\ensuremath{#1}\equiv_{({#3}{#4})}{#2}}
\renewcommand{\fr}[2]{\ensuremath\texttt{fr}({#1}, {#2})}
\newcommand{\stw}[3]{\ensuremath{{#1}{\rightarrow{\!\scalebox{0.7}[0.7]{{$#3$}}}}}{~#2}}

\newcommand{\lsv}[1]{\ensuremath{\texttt{L}_{#1}}}

\newcommand{\synbracket}[1]{\llbracket {#1} \rrbracket}
\newcommand{\VT}[1]{\ensuremath{\mathcal{V}\synbracket{#1}}}
\newcommand{\ETAE}[3]{\ensuremath{\mathcal{E}\synbracket{#1\ \textcolor{blue}{#2}\ \textcolor{red}{#3}}}}
\newcommand{\VG}[1]{G\synbracket{{#1}}}

\newcommand{\carrow}{\Rrightarrow}
\newcommand{\equiva}{\ensuremath{\cong}}
\newcommand{\equivlog}{\ensuremath{\mathrel{\approx_{\text{log}}}}}
\newcommand{\semTypeAE}[6]{#1 \models #2 \equivlog #3 : #4\ \textcolor{blue}{#5}\ \textcolor{red}{#6}}
\newcommand{\ctxEquivAE}[6]{#1 \models #2 \equiva #3 : #4\ \textcolor{blue}{#5}\ \textcolor{red}{#6}}

\newcommand{\vallocs}[1]{\ensuremath{\mathnormal{L}({#1})}}
\newcommand{\varslocs}[2]{\mathnormal{L}_{#2}{(#1)}}
\newcommand{\explocs}[2]{\ensuremath{\mathnormal{L}_{#2}{(\fv{#1})}}}

\newcommand*\recdx[1]{\protect\tikz[baseline=(char.base)]{
            \protect \node[shape=rectangle,draw,inner sep=1.5pt,color=teal, font=\small] (char) {#1};}}
\newcommand{\NUM}[2]{\ensuremath{{\recdx{#1}}}~{#2}}

\newcommand{\ab}[1]{\ensuremath{\textcolor{blue}{#1}}}
\newcommand{\ef}[1]{\ensuremath{\textcolor{red}{#1}}}

\newcommand{\wrt}{{\em w.r.t.}\xspace}

\newcommand{\lambdae}{\ensuremath{\lambda}_{\rm{e}}}
\newcommand{\lambdaa}{\ensuremath{\lambda}_{\rm{a}}}
\newcommand{\lambdaae}{\ensuremath{\lambda}_{\rm{ae}}}

\lstdefinelanguage{DOT}%
{morekeywords={val,new},%
  sensitive,%
  morecomment=[l]//,%
  morecomment=[s]{/*}{*/},%
  morestring=[b]",%
  morestring=[b]',%
  showstringspaces=false%
}[keywords,comments,strings]%

\newlength{\trulemargin}
\newlength{\trulewidth}
\newlength{\srulewidth}
\newenvironment{trules}{$\vspace{0.5em}\ba{p{\trulemargin}@{~}p{\trulewidth}@{~}p{\trulemargin}}}{\ea$}
\newenvironment{srules}{$\vspace{0.5em}\ba{p{\trulemargin}@{~}p{\srulewidth}}}{\ea$}

\newcommand{\ba}{\begin{array}}
\newcommand{\ea}{\end{array}}

\newcommand{\ei}{\end{array}}
\newcommand{\bcases}{\left\{\begin{array}{ll}}
\newcommand{\ecases}{\end{array}\right.}

\newcommand{\eg}{{\em e.g.}\xspace}

\newcommand{\judgement}[2]{{\textsf{\textbf{#1}}} \hfill #2}

\makeatletter%
\begin{document}

\title[Type, Ability, and Effect Systems: Perspectives on Purity, Semantics, and Expressiveness]
{Type, Ability, and Effect Systems: Perspectives on \\Purity, Semantics, and Expressiveness}\thanks{This article is an extended version of \citet{bao2026}.}

\author{Yuyan Bao}
\correspondingauthor
\orcid{0000-0002-3832-3134}
\affiliation{%
  \institution{Augusta University}
  \city{Augusta}
  \country{USA}
}
\email{yubao@augusta.edu}

\author{Tiark Rompf}
\orcid{0000-0002-2068-3238}
\affiliation{%
  \institution{Purdue University}
  \city{West Lafayette}
  \country{USA}
}
\email{tiark@purdue.edu}

\begin{abstract}
Programming benefits from a clear separation between
pure, mathematical computation and impure, effectful interaction
with the world. Existing approaches to enforce this separation
include monads, type-and-effect systems, and capability systems.
All share a tension between precision and usability, and
each one has non-obvious strengths and weaknesses.
This paper aims to raise the bar in assessing such systems. %

First, we propose a semantic definition of purity,
inspired by contextual equivalence, as a baseline for effect soundness
independent of any specific typing discipline.
Second, we propose that expressiveness should be measured by
the degree of completeness, \ie, how many semantically pure terms 
can be typed as pure. 
Using this measure, we focus on minimal meaningful effect and 
capability systems and show that they are incomparable, \ie, 
neither subsumes the other in terms of expressiveness.

Based on this result, we
propose a synthesis and show that \emph{type, ability, and effect systems}
combine their respective strengths while
avoiding their weaknesses.
As part of our formal model, we provide a logical relation 
to facilitate proofs of purity and other
properties for different effect typing disciplines.
 \end{abstract}


\maketitle

\lstMakeShortInline[keywordstyle=,%
                    flexiblecolumns=false,%
                    language=PolyRT,
                    basewidth={0.56em, 0.52em},%
                    mathescape=true,%
                    basicstyle=\small\ttfamily]|

\section{Introduction}\label{sec:intro}

Computer programs bridge two worlds: the ``pure'' world
of mathematical computation and the ``impure'' world
of stateful physical reality.
Some programs are dominantly computational in nature,
with the goal of producing a discrete result upon
termination. The nature of other programs is dominantly
interactive, with the goal of observing and affecting 
state changes in the real world, be it to control
machinery or to interact with users.\footnote{
	While the term ``microcomputer'' has fallen out of fashion,
	the term ``microcontroller'' remains in active use.
}
Many programs combine both aspects, and ultimately,
all programs\footnote{
	Not all programs are ``run'': some are only written
	to illustrate an idea, or to serve as a Curry-Howard proof witness.
} 
are executed on stateful physical hardware.
This creates tensions.

\bfparagraph{Maximizing Mathematical Reasoning}

From the perspective of computation, we often wish to view 
program terms as mathematical formulas, and reason about
them as we do in middle-school algebra. The essence of a 
term behaving in such a ``mathematical'' way is a strong
form of equational reasoning that \emph{identifies}
a term with the result it computes. 
Key are the principles of extensionality and compositionality: 
terms with indistinguishable results should be treated as equivalent,
and it should be possible to replace a subterm with an equivalent 
one without changing the overall result. 
All this crucially depends on a term having 
a single, fixed result, which does not change depending on 
when or how often the term is evaluated. 
In other words, we want to view a program term as a 
mathematical function of its free variables,
consistently producing equivalent outputs on equivalent 
inputs. We can only meaningfully say that $2+3=5$ 
because $2+x$ depends only on the value of $x$ but does 
not depend, \eg, on the phase of the moon.

\bfparagraph{Stateful Physical Reality}

But in the real world, programs often behave ``unmathematically'',
for two different reasons.
The first reason is intentional: often the very purpose of a program
is control, more than computation, \eg, to drive machinery based
on sensor inputs, or to perform tasks based on user interaction.
Observing or affecting a state change in the
physical world is the primary purpose. 
The second reason is accidental: 
even purely computational parts of a program are ultimately executed 
as a stream of machine instructions 
on a stateful microprocessor, which presents the abstraction of ``bits'' 
changing states in ``memory'' on top of some electrical circuitry.
The tower of abstractions on top of the physical reality is colossal,
spanning superscalar processors with out-of-order execution, 
virtual machines, optimizing compilers, 
various storage technologies, %
and unreliable network connections to data centers on 
the other side of the planet.

It takes great care to orchestrate this stack in a way that preserves any 
mathematical illusion. At any level of granularity, a component may present
a pure external interface based on an internal stateful implementation
(effect masking).
Thus, what is pure or impure, and external or internal, always depends 
on \emph{which aspects one is able to observe}.

\bfparagraph{Monads, Effects, Capabilities}

Given the desire to maximize mathematical reasoning and the potential
consequences of accidental impurity where purity is assumed,
programming languages have long sought to enforce purity in parts 
of a program, and to isolate pure from impure parts.

Monads~\cite{DBLP:journals/iandc/Moggi91,
DBLP:conf/popl/Wadler92}, as in Haskell, 
can either encode effects using pure code or isolate 
them at the boundary with the runtime (the IO Monad). 
Effect type systems~\cite{gifford1986integrating,DBLP:journals/iandc/TalpinJ94} 
instead enrich types with annotations that 
track the occurrence of impure operations. 
More recently, there is a resurgence in capability-based 
control of effects~\cite{DBLP:journals/cacm/DennisH66,RobustComposition},
where a type system tracks the flow of values that
grant access to impure operations, without tracking
the operations directly~\cite{osvald2016gentrification,
DBLP:journals/pacmpl/BrachthauserSO20,DBLP:conf/icfem/CraigPGA18,
DBLP:journals/toplas/BoruchGruszeckiOLLB23,
DBLP:conf/ecoop/Gordon19}.

\subsection{Challenges and Limitations}\label{sec:challenges}

Like with any form of typing discipline, all these approaches 
suffer from a tension between precision and usability. 
Monadic encodings add an indirection compared to direct-style
code that can be awkward to deal with. Thus, for pragmatic
reasons, even Haskell supports nontermination and exceptions 
in the ``pure'' fragment of the language (a Haskell function 
of type $\mathbb{N} \to \mathbb{N}$ is not guaranteed to be a 
mathematical function on the natural numbers).
And despite an abundance of research on effect systems,
few are used outside of experimental settings.

\bfparagraph{Effect Polymorphism}

Consider a polymorphic higher-order function |map| 
that applies a given function |f|
to all elements of a given list |ls|.
Ideally, whether the application |map(f)(ls)| is pure or impure should depend on whether |f| is a pure or impure 
function (a partial application |map(f)| should always be pure). 
The first idea is to let |map| accept argument functions with any possible effect:
\begin{lstlisting}
	def map[A, B](f: A => B @eff)(ls: List[A]): List[B] @eff
\end{lstlisting}
This, however, makes every invocation of |map| appear effectful, even when |f| is pure.
For example mapping over a list of integers |lzs| with a pure function, 
\eg, |map(x => x+1)(lzs)|, and with an effectful function, 
\eg, |map(x => println(x))(lzs)|, 
both have the impure type |List[Int] @eff|.
The alternative is to introduce explicit effect polymorphism~\cite{DBLP:conf/popl/LucassenG88}:
\begin{lstlisting}
	def map[A, B, E](f: A => B @eff[E])(ls: List[A]): List[B] @eff[E]
\end{lstlisting}
Now, |E| covers both pure and impure cases,
but the result is a proliferation of type parameters: every higher-order function 
needs to be effect-polymorphic, and in an OOP setting, essentially
every method is a higher-order function parametric in the type of the receiver~\cite{DBLP:conf/scala/OderskyBBLL21}.%
\footnote{One of the widely available effect systems, checked exceptions in Java, demonstrates exactly this problem. Empirically, programmers rarely use the available effect-polymorphism in Java~\cite{DBLP:conf/ecoop/RytzOH12,DBLP:phd/ch/Rytz14,DBLP:journals/toplas/BoruchGruszeckiOLLB23}}

Capabilities have shown great promise for enabling 
effect-polymorphic behavior without explicit parameterization,
so there are high hopes that they may become a viable
solution for modeling effects in mainstream languages
\cite{osvald2016gentrification,DBLP:journals/pacmpl/BrachthauserSO20,
DBLP:journals/toplas/BoruchGruszeckiOLLB23}.
Consider the following definition of the function |map| with capabilities:
\begin{lstlisting}
	def map[A, B](f: (A => B) @cap)(ls: List[A]): List[B]
\end{lstlisting}
The argument type |(A => B) @cap| indicates that |map| accepts a function |f| 
that may have captured capabilities (the list |ls| may not). 
The result type is just type |List[B]|, which is independent of |f|.
The pure expressions are simply those that do not mention any capabilities from the context, %
and thus the purity of |map(f)(ls)| depends on whether |f| has captured any capabilities or not (see \secref{sec:ability_formalization}).

\bfparagraph{Use vs.\ Mention of Resources}

But the convenience of capabilities also comes with trade-offs \wrt 
effect systems, specifically when it comes to distinguishing
actually \emph{using} a capability to cause an effect from just
\emph{mentioning} a capability~\cite{DBLP:conf/ecoop/Gordon19}:
\eg, even the partial application |map(f)| now looks
effectful if |f| mentions any capabilities! 
The following example also illustrates the issue:

\begin{minipage}[t]{0.50\textwidth}
\begin{lstlisting}[aboveskip=-0.5\baselineskip]
def mention(x: Ref[Int] @cap) = x // no effect
\end{lstlisting}
\end{minipage}%
\begin{minipage}[t]{0.5\textwidth}
\begin{lstlisting}[aboveskip=-0.5\baselineskip]
def use(x: Ref[Int] @cap) = { x += 1 } // effect
\end{lstlisting}
\end{minipage}%

\noindent In the above code, the function |mention| does not have an effect, as it merely mentions the variable |x|,
whereas the function |use| actually mutates it. A pure capability system cannot make such a distinction when treating |x| as a store-access capability.
While this trade-off may be acceptable for some applications, it is especially problematic for 
systems that need to track flow-sensitive behavior, \eg, deallocation or ownership transfer~\cite{DBLP:journals/corr/abs-2510-08939}, or systems that encode general typestate transitions~\cite{DBLP:conf/pldi/Liu26} using revocable capabilities. 
The ability to distinguish use vs.\ mention is essential for modeling any form of destructive effect.

\bfparagraph{Effect Masking}

A component may present a pure external interface yet use effects internally:
\begin{lstlisting}
	def sum(xs: List[Int]): Int = { val s = new Ref(0); for (x <- xs) s += x; !s } 
\end{lstlisting}	
In this example, the reference |s| is local to the function |sum| and does not escape from its scope. 
Thus, the effect of mutating |s| is confined within the function, which behaves observationally pure. 
The solution to this challenge is called \emph{effect masking}.
A similar case arises when a function throws and catches an exception locally.
The type system should be able to reflect such cases of purity.

With effect systems, effect masking is typically achieved by adding regions in effect types to track store effects, and allowing one to remove local effects via an effect masking rule; see, \eg, \cite{DBLP:conf/popl/LucassenG88} and \cite{DBLP:conf/popl/Benton0N14}. 
This comes at the price of even higher degrees of parameterization, as higher-order functions now need to be both effect-polymorphic and region-polymorphic~\cite{DBLP:journals/iandc/TalpinJ94}.

For capability systems, in the example above, the function |sum| does not have any free variables, 
and returns a non-capability value. However, the allocation of |s| still needs to be treated as impure,
so the challenge lies in correctly distinguishing scope-local capabilities from those that are captured from 
some outer context. This is hard if there is only a single flag |@cap| to track capabilities,
and generally seems to require more advanced mechanisms~\cite{DBLP:journals/pacmpl/WeiBJBR24,DBLP:journals/pacmpl/XuBO24}.

\bfparagraph{Combining Strengths without Weaknesses?}
In this work, we study different effect typing disciplines formally, and
we present a synthesis, $\lambdaae$, that combines capabilities
and effects in a unified system, using only simple qualifiers equivalent to |@cap| and |@eff|, plus an additional marker for fresh allocations (see below).
Like pure capability systems, $\lambdaae$ supports effect polymorphism without explicit parameters.
Like pure effect systems, it distinguishes use vs.\ mention. 
Like region systems, it supports many cases of effect masking.
The key is in the interplay between the qualifiers: 

\bfparagraph{Effects = Using a Capability}
The core design principle of $\lambdaae$ is to equate effects 
with \emph{using a capability}: the effect qualifier of an expression |t| is interpreted relative to |t|'s
\emph{ambient ability}, \ie, whether |t|'s free variables refer to any capabilities in the context.
The type system admits a strengthening rule (\lemref{lem:hast_strengthen}) to safely downcast a 
typing |t: T @eff| to just |t: T| if the free variables of |t| do not refer to any |@cap| values 
in the environment. %

\bfparagraph{Absence of Effects $\ne$ Purity}
However, this strict notion of effects does not cover all sources of impurity. Specifically,
a term that returns a freshly allocated resource must be treated as impure:
every evaluation returns a different unused address but
does not have an effect in the sense of \emph{using} any (known) resource. 
Special treatment is required, and $\lambdaae$ introduces |@fresh| as
a subqualifier of |@cap| for such situations
(inspired by reachability types~\cite{DBLP:journals/pacmpl/BaoWBJHR21,DBLP:journals/pacmpl/WeiBJBR24}).%
\footnote{Nontermination would also require special treatment (see also \secref{sec:purity_intuition}); however,
all $\lambdaae$ programs terminate.}

\bfparagraph{Effect Polymorphism, Revisited}

Consider again the case of |map|, which is now typed as follows:
\begin{lstlisting}
	def map[A, B](f: (A => B @eff) @cap)(ls: List[A]): List[B] @eff
\end{lstlisting}

Even though |map| has a latent effect, an application |map(f)(ls)| will be pure when the argument
function |f| and list |ls| do not have any captured capabilities. Through the strengthening property, 
|map(f)(ls): List[B] @eff| can be downcast to |map(f)(ls): List[B]|, ignoring the latent effect of |map| 
(see \secref{sec:tae} for the details). 
Thus, we achieve the same behavior as in the system with only |@cap|, which, in this case, we considered 
preferable to the system with only |@eff|. In other cases, such as just |map(f)| or the example of |mention| and |use| above, 
we obtain the desired behavior of |@eff|.

\bfparagraph{Effect Masking, Revisited}
For effect masking, the distinction between |@fresh| and |@cap| allows 
us to distinguish between newly allocated capabilities and 
those from the surrounding context. Consider the following example:
\begin{lstlisting}
	def inc(x: Ref[Int] @cap): Bool @eff = { x += 1; true } 
\end{lstlisting}
When applying |inc| to a freshly allocated reference as in |inc(new Ref(0))|, we again
have a term without ambient ability (no |@cap| values referenced from the context),
so the strengthening property allows us to downcast |inc(new Ref(0)): Bool @eff| 
to just |inc(new Ref(0)): Bool|,
safely ignoring |inc|'s latent effect as changes to a fresh resource cannot be observed.

\subsection{Contributions}

How should we compare and evaluate different effect typing disciplines?
While there are well-known results connecting monads and 
effects~\cite{DBLP:journals/tocl/WadlerT03,DBLP:conf/popl/Filinski10}, 
less is known about the relationships between different effect and capability systems. 
The key goal of this paper is to raise the level of rigor in this area and establish a framework for comparing such diverse typing disciplines. To achieve this, we must first establish a common semantic foundation. Rather than relying solely on syntactic criteria or language-specific features, we propose a semantic soundness approach based on a contextual notion of purity, inspired by contextual equivalence. 

Using this framework, we can compare the expressiveness of different typing disciplines by measuring their degree of completeness in characterizing semantically pure terms. This allows us to assess how well each system captures the intended notion of purity and to analyze their respective strengths and weaknesses. We illustrate this approach by formalizing and comparing canonical binary effect and capability systems, $\lambdae$ resp. $\lambdaa$, demonstrating that they are incomparable in terms of expressiveness. 
This means that for each system, there exist programs that can be typed as pure, which cannot be typed as pure in the other system. So neither system subsumes the other in its ability to identify semantically pure programs.
We then present our synthesis, $\lambdaae$, 
with the goal
to address the challenges identified above: effect polymorphism, masking, use vs.\ mention.

To facilitate further research and practical reasoning about purity, we present a logical relation that enables formal proofs of purity across different typing disciplines. This relation is designed to support both capabilities and effects, supporting a variety of type system designs.

In summary, the main contributions of this paper are:
\begin{itemize}[left=2ex]
	\item A framework for measuring the expressiveness of effect typing disciplines based on their ability to characterize semantically pure terms as pure (\secref{sec:semantics}).
	\item A semantic definition of purity, independent of any particular type system, inspired by contextual equivalence (\secref{sec:purity}).
	\item Formalization and comparison of canonical binary effect and capability systems, $\lambdae$ resp. $\lambdaa$, demonstrating that they have incomparable expressiveness (\Cref{sec:effects,sec:ability}).
	\item A synthesis of effect and capability systems, $\lambdaae$, combining their respective strengths to address identified weaknesses, and shown to subsume both earlier systems (\secref{sec:tae}).
	\item A general logical relation suitable for proving purity, applicable to a range of typing disciplines and language designs (\secref{sec:lr}).
\end{itemize}

We discuss possible extensions and related work in \secref{sec:diss} resp. Section~\ref{sec:related}, and offer concluding
remarks in Section~\ref{sec:conc}. Mechanized proofs and accompanying
developments are available online at~\url{https://github.com/tiarkrompf/reachability}.

\section{Semantics of Effects and Purity}
\label{sec:semantics}

We begin by establishing some intuition behind our framework for reasoning about effects and purity in programming languages. This framework will allow us to compare different effect typing disciplines and their expressiveness.

\subsection{Expressiveness of Effect Typing Disciplines}
\label{sec:expressiveness}

Felleisen's notion of macro-expressibility~\cite{DBLP:journals/scp/Felleisen91} provides a formal foundation for understanding the expressive power of programming languages. However, when it comes to type systems, the meaning of expressiveness is less well established.

One reasonable and, we believe, common interpretation is that a more expressive type system can type more programs, such as in the case of System F vs.\ simply-typed  $\lambda$-calculus~\cite{samthExpressiveness11}. This of course assumes that we consider only type systems that are sound, as unsound type systems can trivially type all programs. Thus, expressiveness can be understood as the degree of completeness of a type system, \ie, the proportion of all semantically well-typed programs (in the sense of ``well-typed programs don't go wrong') that are well-typed syntactically.
This common perspective also applies to scenarios where one assumes a fixed set of types but considers extensions to the term language~\cite{DBLP:books/sp/LongleyN15,DBLP:journals/lmcs/Longley18,DBLP:journals/jfp/HillerstromLL24}.

But can we apply the same measure to effect typing disciplines? If we only consider the number of typeable terms, the metric becomes trivial: since effects refine an existing type system, we could simply type all terms as impure. Clearly, this does not capture the essence of expressiveness for effect systems. The key realization, of course, is that for an effect systems to be sound, it should reflect stronger semantic properties than just ``well-typed programs don't go wrong''. These  properties may depend on the specific effect system so they are hard to capture generically. More disturbingly, they are often not precisely defined, especially when effects are modeled purely syntactically: unlike with simple type systems, there is no notion of effects being preserved under reduction from terms to values, and there is no canonical equivalent to ``getting stuck'' for effect typing violations.%
\footnote{Systems that delegate the interpretation of effects to \emph{effect handlers}~\cite{DBLP:journals/corr/PlotkinP13} do have a notion of getting stuck when an effect is performed without a corresponding handler.} 
Any attempt at artificially forcing stuckness in a reduction semantics may overlook possible error cases, so we believe that a more semantic approach is 
indispensable for a rigorous treatment.

Irrespective of the precise semantic effect soundness properties of any specific effect system, we propose that any meaningful effect system should at least satisfy the property that syntactically pure terms are also semantically pure.
Thus, a natural way to measure the expressiveness of different effect systems is by asking: how many semantically pure terms can the system type as pure? This leads to the next question: what exactly does it mean for a term to be semantically pure?

\subsection{Intuitive Definitions of Purity}\label{sec:purity_intuition}

Perhaps surprisingly, there does not seem to be a universally agreed-upon definition of what it means for a term to be pure. However, there are several common intuitions that we can build upon. We propose that the following three intuitions are equivalent and capture the essence of purity: 

\begin{enumerate}
\item\textbf{Pure terms permit reasoning ``as in Math''.}
This alludes to a strong form of equational reasoning that \emph{identifies} a term with its result value.\footnote{It is of course possible to reason equationally about impure terms by means of their \emph{denotations}; however, those denotations are not just \emph{values}, but include, \eg, state transformations. Such terms do not exist in elementary Mathematics.} 

\item\textbf{Pure terms are mathematical functions of their free variables.}
This stresses the principles of extensionality and compositionality, as a basis for intuitive context-free rewriting.

\item\textbf{Pure terms terminate, have deterministic results, and evaluate without effects.}
Effects include (1) primary effects, \ie, external state changes such as IO; 
(2) observable side effects, \ie, internal state changes that impact the result of evaluating any \textit{other term}.

\end{enumerate}

Among the possible effects of a term, externally observable effects, \eg, IO (file or network reads and writes), are the easiest to capture with a type system, hence we will mostly disregard them. The more interesting case is that of observable side effects, \ie, internal state changes that impact the result of evaluating other terms.
Figure~\ref{fig:intuitionExamples} shows examples of pure and impure terms,
including cases of effect polymorphism, effect masking, and use vs.\ mention
situations.

\begin{figure}\small
\begin{mdframed}[leftmargin=2pt, rightmargin=2pt]
$$\begin{array}{p{2cm}p{3cm}p{3cm}p{2cm}}
\textsc{(Abs:Any)} & 
$\tabs t$ & 
\text{Function value} & \text{Pure} \\
\textsc{(Diverge)} & 
$(\lambda x. \tapp x x)\ (\lambda x. \tapp x x)$ & 
\text{Nontermination} & \text{Impure} \\
\textsc{(Alloc)} &
$\tref t$ &
\text{Resource allocation} & \text{Impure} \\
\textsc{(Use:Read)} &
$\tget a$ &
\text{Resource use} & \text{Impure} \\
\textsc{(Use:Write)} &
$\tput a t$ &
\text{Resource use} & \text{Impure} \\
\textsc{(Mention)} &
$\tlet x a t$ &
\text{Resource mention} & \text{Pure iff $t$ pure} \\
\textsc{(Mask:Read)} &
$\tlet x {\tref t} {\tget x}$ &
\text{Masked resource use} & \text{Pure iff $t$ pure} \\
\textsc{(Mask:Write)} &
$\tlet x {\tref t} {\tput x t'}$ &
\text{Masked resource use} & \text{Pure iff $t,t'$ pure} \\
\end{array}$$
\end{mdframed}
\caption{Examples illustrating intuitively pure and impure terms
($a$ refers to a mutable reference).
}
\label{fig:intuitionExamples}
\end{figure}

\begin{figure}\small
\begin{mdframed}[leftmargin=2pt, rightmargin=2pt]
$$\begin{array}{p{2cm}p{3cm}p{3cm}p{2cm}}
\textsc{(Abs:Id)} & 
$\tabs x$ & 
\text{Pure function} & \text{Pure ability} \\
\textsc{(Abs:Diverge)} & 
$\tabs \omega$ & 
\text{Impure function} & \text{Impure ability} \\
\textsc{(Abs:Alloc)} & 
$\tabs {\tref x}$ & 
\text{Impure function} & \text{Impure ability} \\
\textsc{(Abs:Leak)} & 
$\tabs a$ & 
\text{Pure function} & \text{Impure ability} \\
\textsc{(Abs:Use)} & 
$\tabs {\tget a}$ & 
\text{Impure function} & \text{Impure ability} \\
\textsc{(Abs:UseArg)} & 
$\tabs {\tget x}$ & 
\text{Impure function} & \text{Pure ability} \\
\textsc{(Abs:Mention)} & 
$\tabs {\tlet x a \ttrue}$ & 
\text{Pure function} & \text{Pure ability} \\
\textsc{(Abs:Capture)} & 
$\tabs {\lambda y.\ {\tget a}}$ & 
\text{Pure function} & \text{Impure ability} \\
\textsc{(Abs:Poly)} & 
$\lambda f.\ \lambda x.\ {\tapp f x}$ & 
\text{Impure function} & \text{Pure ability} \\
\end{array}$$
\end{mdframed}
\caption{Examples illustrating intuitively pure and impure 
functions (all are pure terms here, but could also be
results of impure computation),
and the concept 
of \emph{ability} %
of a value: namely, 
to be the source of impure behavior when put in a 
specific context. Values with impure ability are
intuitively understood as \emph{capabilities}.}
\label{fig:intuitionExamples2}
\end{figure}

\subsection{Pure Functions, Pure Ability}

To state whether a function application is pure, we also need
the concept of a \emph{pure function}. Focusing on its
extensional behavior, we regard a term $f$ as a pure function iff,
when $f$ is applied to an argument $t$, 
then the overall application term $\tapp f t$ is pure,
disregarding any impurity caused by evaluating the subterms $f$ or $t$ themselves
(\ie, only considering impurity caused by the application itself).
Figure~\ref{fig:intuitionExamples2} shows examples of pure and 
impure functions.

To think about effects in terms of using capabilities, we
introduce the notion of a term's \emph{ability}
(following terminology introduced by \citet{DBLP:conf/popl/LindleyMM17}).
An impure function may still have pure ability
in the sense that it can't perform an effect on its own, but
it needs another object, a \emph{capability}, to cause this effect.
Focusing again on extensional aspects, we regard a term $f$ 
as a pure-ability function iff,
when $f$ is applied to a pure-ability argument $t$, 
then the overall application term $\tapp f t$ is pure
(disregarding impurity caused by evaluating $f$ or $t$),
and the result is also a pure-ability value. 
In addition to pure-ability functions, primitives
also count as pure-ability values.

Intensionally, ability describes whether a term's result has 
captured any effect-inducing capabilities, such as mutable references.
For example, $\tabs {\tget x}$ is not a pure function because it
dereferences its argument $x$, but it has pure ability
because the reference needs to be provided externally, it
has not been captured by the function.
On the other hand, $\lambda f.\ \tabs {\tapp f x}$ is not a pure 
function because when applied to the argument $\tabs {\tget a}$,
a pure term, the application is impure. 
Likewise, a pure function may also have impure ability:
$\tabs {a}$ is a pure function, but it returns a 
reference it has captured. %
Finally, $\tabs {\tref x}$ is an impure function, and it
also has impure ability. It does not capture the ref it
returns, but it still returns a value capable of inducing
an effect elsewhere.
To distinguish these cases, we will refer to $\tabs {a}$
as having \emph{ambient} ability and $\tabs {\tref x}$
as having \emph{innate} ability.

\section{A Semantic Definition of Purity}
\label{sec:purity}

In the following, we propose a semantic notion of purity that matches the intuition outlined above. 
The definition is so simple and straightforward that we are almost certainly not the first to operate with this definition in mind. %
Specifically, our definition of pure terms matches that of algebraic values given by \citet{DBLP:conf/fossacs/Fuhrmann02} and closely corresponds to that of thunkable expressions/morphism by \citet{Hayothesis,DBLP:journals/entcs/Fuhrmann99}.
However, we believe we are the first to propose this property as a general definition of purity, rather than for a somewhat narrower purpose such as thunkability.

\subsection{Formalizing A Notion of Observational Purity}

\subsubsection{Syntax and Semantics}
To ground this discussion, we fix some minimal requirements on the
program syntax and semantics.
We consider the syntax consisting of terms $t$, values $v$, and states $\sigma$, along with 
value environments $H$, which map variables to values: 
$$
\small
\ba{rcll}
b &:=& \ttrue\ \mid\ \tfalse\ \mid\ ... & \text{Constants} \\
t &:=& b\ \mid\ \tvar x\ \mid\ \tlet x t t\ \mid\ ... \qquad & \text{Terms} \\
v &:=& b\ \mid\ ... & \text{Values} \\
\sigma &:=& ... & \text{State} \\
H &:=& \varnothing \mid H \extends (x, v)  & \text{Value Environments} \\
\ea
$$
We intentionally leave the precise structure of terms, values, and state underspecified at this stage. 
Specifically, we do not assume any specific notion of functions, and we do not prescribe any general evaluation order (call-by-value, call-by-name, call-by-need).
However, we do assume that the term language supports sequencing via let-bindings
whose binding semantics are strict, \ie, fully evaluate the right-hand side before the body.
Note here the connection with Moggi's computational $\lambda$-calculus~\cite{DBLP:conf/lics/Moggi89} (and specifically the computational metalanguage), 
which introduced monads as a model of effectful computation. 
Among the set of values, we assume a distinguished subset of atomic primitives (\eg, integers and booleans), 
which serve as effect-free observables.

The operational semantics defines evaluation as a relation of the 
form
$\config{H}{\sigma}{t} \eval \pconfig{\sigma'}{v}$. 
This means that evaluation 
of term $t$ with value environment $H$ and pre-state $\sigma$ 
terminates with result value $v$ and post-state $\sigma'$.
We assume that evaluation itself is deterministic, \ie, 
there exists no more than a single $v$ and $\sigma'$ for any given $t$, $H$, and $\sigma$.
If evaluation is undefined in the case of an error or divergence, and 
thus no $v$ and $\sigma'$ exists for a given $t$, $H$, and $\sigma$, we also write
$\config{H}{\sigma}{t} \Uparrow$.

Constants evaluate to themselves, and the evaluation of $\texttt{let}$
expressions is strict. As mentioned above, this is important to guarantee 
proper sequencing.
Later, we will formally extend the language with $\lambda$-abstractions 
and mutable references (in \secref{sec:concreteSemantics}). Extensions with IO, exceptions, etc., are also
possible, but we prefer to keep our core definitions independent of 
any such specific features.

In the following, we will largely keep the environment $H$ and 
state $\sigma$ implicit. So when clear from the context, we will
just write ${t} \Uparrow$ and ${t} \Downarrow v$ to mean that
$t$ evaluates in its associated $H$ and $\sigma$.

\subsubsection{Observational Equivalence}\label{sec:obs_equiv}

Our notion of \emph{observational} (or \emph{contextual}) purity is inspired crucially by the notion of observational (or contextual) \emph{equivalence}, typically attributed to Morris' 1969 PhD thesis (there called \emph{extensional equivalence})~\cite{morris1968lambda}. 
This turned out to be a key milestone in the ability to reason about programs equationally, 
as the straightforward identification of terms that evaluate to \emph{identical} values is too restrictive.
One could consider equivalence of result values modulo some conversion rules, such as those in Morris' thesis~\cite{morris1968lambda}.
However, in a language with references, this is not straightforward. For example, one does not want to make any assumptions that allocation terms such as $\tref 0$ and $\tref 0$ would reduce to identical store locations.

Instead, one sets up a weaker form of equivalence first
that only considers termination behavior and atomic result values
such as integers and Booleans\footnote{%
Often the requirement to agree on atomic result values is dropped (\ie, requiring just the same termination behavior) but we prefer to include this condition, as it is necessary for terminating languages.}
(for the definitions below, we assume evaluation 
in two equivalent pre-states $\sigma_1$ and $\sigma_2$, for a suitable notion of state equivalence).

\begin{definition}[Kleene Equivalence]\label{def:opEquiv}
    Two terms $t_1$ and $t_2$ are Kleene equivalent $t_1 \simeq t_2$ iff both have the same termination behavior and they agree on
    any atomic result values $b$:
\[
\small 
{t_1} \Uparrow\ \  \Leftrightarrow \ {t_2} \Uparrow \qquad \wedge \qquad 
{t_1} \Downarrow {b} \ \Leftrightarrow \   {t_2} \Downarrow {b}
\]
\end{definition}

Then, the key idea is to treat all those terms as equivalent for which the \emph{program itself} cannot observe any difference. So, no context should exist that would expose any difference in behavior:

\begin{definition}[Observational Equivalence]\label{def:oeEquiv} Two terms $t_1$ and $t_2$ are observationally equivalent $t_1 \cong t_2$ iff for all contexts $C$:
\[
\small 
C[t_1] \simeq C[t_2] 
\]
\end{definition}

\subsubsection{Observational Purity}
By analogy with observational equivalence, we want to express that pure expressions are those that do nothing else (observably) than computing their results. So, no context should exist that would expose any difference in behavior between the term itself, possibly evaluated an arbitrary number of times (including zero) by the context, and evaluating the term once outside the context and then reusing its result bound to a variable (with strict binding semantics):

\begin{definition}[Observational Purity]\label{def:opure} A term $t$ is observationally pure iff for all contexts $C$ which do not bind free variables of $t$, and fresh variables $x$:
\[
\small 
\tlet {x} {t}{C[x]} \cong C[t]
\]
\end{definition}

In a language that includes the call-by-value $\lambda$ calculus, the left-hand side $\tlet {x} {t}{C[x]}$ can be encoded as $\lambda$ application $\tapp {(\lambda x. C[x])} t$. This suggests a nice symmetric interpretation of pure terms as (a) invariant under $\eta$-expansion of any context $C$ into a $\lambda$ abstraction, and (b) invariant under $\beta$-reduction in any application position. 
The latter interpretation is consistent with Sabry's definition of purity~\cite{DBLP:journals/jfp/Sabry98}, modulo
our additional requirement for termination.

In terms of terminology, we consider a term observationally impure if it is not observationally pure. When clear from the context, we will drop the ``observationally'' qualifier.

\subsection{Observational Purity: From Untyped to Typed Contexts}\label{sec:refine}

Even though observational equivalence is a purely semantic
notion, there are important differences when considering untyped vs.\ typed versions of the
same language. The reason is that in a typed setting, we will not consider
any ill-typed contexts as eligible observers in the definition of equivalence.
Thus, in a typed setting, the equational theory is generally richer, with more
terms being considered equal. The following example illustrates this point. 

Consider the term $\lambda x. \neg(\neg x)$, where $\neg x$ denotes Boolean negation of $x$. We might expect this term to be equivalent
to $\lambda x. x$. 
As we will see, this is true in a typed setting, but not in an
untyped setting.
According to \defref{def:oeEquiv}, we are faced with the obligation to show $C[\lambda x. \neg(\neg x)] \simeq C[\lambda x. x]$, for all contexts $C$.
In a typed setting, both terms will be typed as $\TBoolA \to \TBoolA$, and we will only
consider contexts $C$ that accept a term of type $\TBoolA \to \TBoolA$ placed in the hole, ruling out any type errors at runtime.
In an untyped setting, however, we have no such restrictions, and it is easy to
find a context $C$ for which the equivalence does not hold. 
For example, the context $C$ may apply the function to a non-Boolean value, such that
the left-hand side will fail with an error whereas the right-hand side always evaluates successfully. 

As is standard, we refine the definition of observational equivalence with a typed environment. We write $C:(\Gamma; T) \carrow (\qempty; \TBoolA)$ to mean that the context $C$ is a closed program, and has type $\TBoolA$,
with a hole that can be filled with any program $t$, such that $\G \ts t: T$.
In doing so, we assume a suitable type system which may or may not include
effect tracking. %
We will consider concrete type systems in 
Sections~\ref{sec:effects} to ~\ref{sec:tae}, but the definitions below are independent of any
specific type system.

\begin{definition}[Typed Observational Equivalence]\label{def:opEquivty} 
    Terms $t_1$ and $t_2$ are observationally equivalent at type $T$, written $\G \models t_1 \cong t_2 : T$, 
    if $\G \ts t_1 : T$ and $\G \ts t_2: T$, and for all $C: (\Gamma; T) \carrow (\qempty; \TBoolA)$:
\[
\small 
    C[t_1] \simeq C[t_2]
 \]
 \end{definition}

As before, we define obs. purity in terms of obs. equivalence, but this time in a typed setting:

\begin{definition}[Typed Observational Purity]\label{def:opurety} 
    A term $t$ is observationally pure at type $T$, if $\G \ts t: T$, and 
    for all $C:(\Gamma;T) \carrow (\Gamma'; T')$ and $x$ fresh:
\[
\small
\setlength{\abovedisplayskip}{0pt}
\setlength{\belowdisplayskip}{0pt}
    \Gamma' \models  \tlet {x} {t}{C[x]} \cong C[t] : T'
\]
\end{definition}

We will consider a term $t$ to be pure, if it is observationally pure at all types $T$ for which $\G \ts t: T$ for some $\G$. Thus, an ill-typed term will be considered pure since its impurity can never be observed.

Going back to our example, the term $\lambda x. \neg(\neg x)$ cannot be  
considered observationally pure in an untyped setting (\defref{def:opure}) because
an ill-typed context may observe a type error. But the same term will be considered
pure in a typed setting (\defref{def:opurety}) as no type error can occur.

Thus, the theory of typed purity is generally richer than the theory of
untyped purity, which always has to consider impurity due to errors.
With the goal of comparing effect typing disciplines, the rest of the paper
will mainly consider typed settings, although we will sometimes
consider purity in untyped contexts as well.

\subsection{Discussion}
\label{sec:concreteSemantics}

\begin{figure}[t]\small
    \begin{mdframed}[leftmargin=2pt, rightmargin=2pt]
    \typicallabel{T-False-E}
    \judgement{Syntax}{$\BOX{t ::= ...}$}\vspace{-2ex}
    \begin{minipage}[t]{\linewidth}
        \[\begin{array}{l@{\quad}l@{\quad}l@{\quad}l}
                b       & :=  & \ttrue \mid \tfalse                                                           & \text{Constants}       \\
                t       & ::= & b \mid x \mid \tabs{t} \mid \tapp{t_1}{t_2} \mid \tref{t} \mid\ \tget{t}  \mid \tput{t_1}{t_2} \mid \tseqb{t_1}{t_2}  & \text{Terms}         \\
                v       & ::= & b \mid \ell \mid \cll{H}{\tabs{t}}                      & \text{Values}              \\
                H       & ::= & \varnothing \mid H \extends (x, v)                                                           & \text{Value Environment}   \\
                \sigma  & :=  & \varnothing \mid \sigma \extends (\ell, v)                                                             & \text{Store} \\
            \end{array}\]\\
        \end{minipage}%
    \vspace{-4ex}\\
    \judgement{Operational Semantics}{$\BOX{\config{H}{\sigma}{t} \eval \pconfig{\sigma'}{v}}$}\vspace{-4ex}
    \begin{multicols}{2}
    
        \infax[e-cst]{
            \config{H}{\sigma}{b} \, \eval \, \pconfig{\sigma}{b}
        }

        \infrule[e-bin]
        {\config{H}{\sigma}{t_1} \, \eval \, \pconfig{\sigma'}{b_1} \\
            \config{H}{\sigma'}{t_2} \, \eval \, \pconfig{\sigma''}{b_2}}
        {\config{H}{\sigma}{\tseqb{t_1}{t_2}} \, \eval \, \pconfig{\sigma''}{b_1 \synbracket{\nand} b_2}}

        \infrule[e-ref]{
            \config{H}{\sigma}{t} \, \eval \, \pconfig{\sigma'}{v}  \\  \ell = \|dom(\sigma)\| %
        }{
            \config{H}{\sigma}{\tref{t}}\, \eval \, \pconfig{\sigma'\extends(\ell, v)}{\ell}
        }
 
        \infrule[e-get]{
            \config{H}{\sigma}{t} \, \eval \, \pconfig{\sigma'}{\ell}  \quad \sigma'(\ell) = v
        }{
            \config{H}{\sigma}{\tget{t}} \, \eval \, \pconfig{\sigma'}{v}
        }

        \columnbreak
       
        \infrule[e-var]{
            H(x)  =  v
        }{
            \config{H}{\sigma}{x} \, \eval \, \pconfig{\sigma}{v}
        }

        \infax[e-abs]{
            \config{H}{\sigma}{\tabs{t}} \, \eval \, \pconfig{\sigma}{\cll{H}{\tabs{t}}}
        }   
        
         \infrule[e-app]{
            \config{H}{\sigma}{t_1} \, \eval \, \pconfig{\sigma'} {\cll{H'}{\tabs{t}}}\\
            \config{H}{\sigma'}{t_2} \, \eval \, \pconfig{\sigma''}{v_1} \\
            \config{H' \extends(x, v_1)}{\sigma''}{t} \, \eval \, \pconfig{\sigma'''}{v_2}
        }{
            \config{H}{\sigma}{\tapp{t_1}{t_2}} \, \eval \, \pconfig{\sigma'''}{v_2}
        }

        \infrule[e-put]{
        \config{H}{\sigma}{t_1} \, \eval \, \pconfig{\sigma'}{\ell} \\  \config{H}{\sigma'}{t_2} \eval \pconfig{\sigma''}{v}
        }{
        \config{H}{\sigma}{\tput{t_1}{t_2}} \, \eval \, \pconfig{\sigma''[\ell \mapsto v]}{\ttrue}
        }

    \end{multicols}

    \end{mdframed}
    
    \caption{Syntax and Semantics. %
    }
    \label{fig:syntax}
\end{figure}
     
We now fix a concrete language, 
with which we examine our definition of purity with a few examples.
\figref{fig:syntax} defines the language syntax and semantics, 
which extends STLC with mutable references and Boolean expressions. 
The definition is standard and follows~\citet{DBLP:books/daglib/0005958}.
We use a big step semantics with a value environment $H$ and a store $\sigma$,
where $H$ is a partial function that maps variables to values, 
and 
$\sigma$ is a partial function that maps locations to values.
As before, we write $\config{H}{\sigma}{t} \eval \pconfig{\sigma'}{v}$ to mean that 
term $t$ evaluates to value $v$, with a store transition from $\sigma$ to $\sigma'$.
In the semantics, we write $\synbracket{\nand}$ to denote the interpretation of 
the syntactic Boolean $\text{NAND}$ operator $\nand$ as its corresponding meta-level operator.
In the rule \textsc{e-ref}, we write $\|dom(\sigma)\|$ to mean the size of the store.
Other definitions of the semantics are standard.

Following~\cite{Jeremy,DBLP:conf/popl/ErnstOC06,DBLP:conf/popl/AminR17,DBLP:conf/ecoop/WangR17}, 
we extend the big-step semantics $\eval$ to a total evaluation function by adding a numeric fuel value and explicit \text{timeout} and \text{Error} results. 
This semantic definition leads to strong soundness results in \secref{sec:lr}.

\bfparagraph{Nontermination}

A nonterminating expression is \emph{impure}, as expected. 
Consider an example: $\omega = (\lambda x. \tapp x x)\ (\lambda x. \tapp x x)$. 
The context $C = (\tlet y {(\tabs \square)} \ttrue)$ distinguishes between the term $\omega$ and $\omega$ bound to a variable: $\tlet x \omega C[x]$ diverges, but $C[\omega] = (\tlet y {(\tabs \omega)} \ttrue)$ does not.
With a level of abstraction, however, we obtain a \emph{pure} expression: $\tabs \omega$ behaves the same in all contexts for both $\tlet x {\tabs \omega} C[x]$ and $C[\tabs \omega]$. 
However, this expression is not a \emph{pure function}, because there are impure applications, and in fact any application will be impure (e.g., $\tapp {(\lambda x. \omega)} \ttrue$).
Thus, it is also not a \emph{pure-ability function}.
However, if we only consider simply-typed $\lambda$-calculus, $\omega$ is not well typed.
Thus, divergence cannot be observed, and we may treat $\omega$ as pure.

\bfparagraph{Resource Allocation}

An allocation is \emph{impure}.
Consider the expression $\tlet x {\tref t} C[x]$. 
This means that only a single location is allocated and then reused through the variable $x$; 
whereas $C[\tref t]$ means that potentially many locations (or none) are allocated. 
It is easy to craft a context to observe this difference.
For example, $\tlet x {\tref 0} \tput{x} {{\tget x}+1}; \tget{x}$ yields value 1, 
reflecting the single reference being updated.
In contrast, $\tput{(\tref 0)}{{\tget {(\tref 0)}} + 1}; \tget(\tref 0)$ repeatedly allocates fresh locations and thus yields value 0.

\bfparagraph{Resource Mentioning and Usage}
Merely referencing a variable bound in the environment is pure, but dereferencing it is not.
Consider the expression $\tlet x a C[x]$, where $a$ is a mutable reference in the environment.
Since both $x$ and $a$ are variables, $\tlet x a C[x]$ and $C[a]$ must behave the same. However, if we consider $\tlet x {\tget a} C[x]$, then the two expressions may behave differently. Thus, we achieve a proper distinction between \emph{mentioning} a resource and \emph{using} it. This distinction also extends from pure terms to pure functions.

\bfparagraph{Effect Masking}
The expression $\tlet x {\tref t} {\tget x}$ is \emph{pure}, 
despite the internal use of a mutable reference. 
It remains pure if we insert a ``proper'' side effect, such as a write operation on $x$,
\eg, $\tlet x {\tref t} \tput{x} {{\tget x}+1}; \tget{x}$. 
This demonstrates the strength of the notion of observational purity: if store changes or other side effects cannot be observed, it is as if they did not exist.

\section{Effect Type Systems ($\lambdae$)}
\label{sec:effects}

\begin{figure}[t]\small
    \begin{mdframed}[leftmargin=2pt, rightmargin=2pt]
    \typicallabel{T-False-E}
    \judgement{Type \& Effects}{$\BOX{\TE::=...}\ \BOX{e::=...}$} \\\vspace{-1.5ex}
  \[
    \begin{array}{l@{\quad}l@{\quad}l@{\quad}l@{\quad}l@{\quad}l@{\quad}l@{\quad}l@{\quad}l@{\quad}l@{\quad}l@{\quad}l}
      \TE & ::= & \TBoolE \mid  \TRefE \mid \TFunE{\TE_1} {\TE_2} {e_2}  & & \textcolor{red}{e} & := & \textcolor{red}{\bfalse} \mid \textcolor{red}{\btrue}  & & \textcolor{red}{e \rhd e'} & := & \textcolor{red}{e \vee e'}  
    \end{array}
  \]  
  \vspace{-4ex}\\
    \judgement{Term Typing}{$\BOX{\hasTypeE \GE t \TE e}$}\vspace{-4ex}
    
    \begin{multicols}{2}
    
    \infax[T-Cst-E]{
      \hasTypeE \GE {b} \TBoolE \bfalse
    }
    
    \infrule[T-Bin-E]{
      \hasTypeE \GE {t_1} \TBoolE {e_1} \\
      \hasTypeE \GE {t_2} \TBoolE {e_2}
    }{
      \hasTypeE \GE {\tseqb{t_1}{t_2} }\TBoolE {e_1 \rhd e_2}
    }

    \infrule[T-Ref-E]{
      \hasTypeE \GE t \TBoolE e
    }{
      \hasTypeE \GE {\tref t} \TRefE \btrue
    }
    
    \infrule[T-Get-E]{
      \hasTypeE \GE t \TRefE e
    }{
      \hasTypeE \GE {\tget t} \TBoolE \btrue
    }
    
    \infrule[T-Put-E]{
      \hasTypeE \GE {t_1} \TRefE {e_1} \\
      \hasTypeE \GE {t_2} \TBoolE {e_2}
    }{
      \hasTypeE \GE {\tput {t_1} {t_2}} \TBoolE \btrueE
    }
    
    \columnbreak
    
    \infrule[T-Var-E]{
      \indexrTypeE \GE x \TE
    }{
      \hasTypeE \GE {\tvar x} \TE \bfalse
    }
    
    \infrule[T-Abs-E]{
      \hasTypeE {\GE, x : \TE_1} t {\TE_2} {e_2}
    }{
      \hasTypeE \GE {\tabs t} {(\TFunE {\TE_1} {\TE_2} {e_2})} \bfalse
    }
    
    \infrule[T-App-E]{
      \hasTypeE \GE {t_1} {(\TFunE {\TE_1} {\TE_2} {e_2})} {e_f} \\
      \hasTypeE \GE {t_2} {\TE_1} {e_1}
    }{
      \hasTypeE \GE {\tapp {t_1} {t_2}} {\TE_2} {e_f \rhd e_1 \rhd e_2}
    }
    
    \infrule[T-Sub-E]{
      \hasTypeE \GE t {\TE_1} {e_1} \andalso
      \subtypeTE{\TE_1}{\TE_2} \\ 
      \subtypeEE{e_1}{e_2}
    }{
      \hasTypeE \GE t {\TE_2} {e_2}
    }
    
    \end{multicols}
    
    \end{mdframed}
    
    \caption{Effect Type System $\lambdae$.}
    \label{fig:has-type-e}
    \end{figure} %
\begin{figure}[t]\small
    \begin{mdframed}
    \typicallabel{T-False-A}

    \judgement{Subtyping}{$\BOX{\subtypeTE{\TE_1}{\TE_2}}\ \BOX{\subtypeEE{e_1}{e_2}}$}\vspace{-2ex}
    \begin{multicols}{2}
      \infax[S-Cst-E] { 
        \subtypeTE{\TBoolE}{\TBoolE}
      }

      \infax[S-Tref-E]{ 
        \subtypeTE{\TRefE}{\TRefE} 
      }

      \columnbreak

      \infrule[S-Abs-E]{
         \subtypeTE{\TE_3}{\TE_1} \andalso
         \subtypeTE{\TE_2}{\TE_4} \\
         \subtypeEE{e_1}{e_2} 
      }{
        \subtypeTE{\TFunE{\TE_1}{\TE_2}{e_1}}{\TFunE{\TE_3}{\TE_4}{e_2}}
      }
      
    \end{multicols}
    \begin{multicols}{3}
    \infax[S-E] {
      \subtypeEE{\bfalse}{\btrue}
    }

    \infax[S-$\bfalse$-E]{
      \subtypeEE{\bfalse}{\bfalse}
    }

    \infax[S-$\btrue$-E]{
      \subtypeEE{\btrue}{\btrue}
    }

    \end{multicols}
\end{mdframed}
\caption{Effect Subtyping Rules $\lambdae$.} 
\label{fig:sub-type-e}
\end{figure} 
We now consider our semantic notion of purity in the context of type systems that distinguish pure and impure terms. 
The classic approach to do so is via a type and effect system.
As basis, we consider a version of simply-typed $\lambda$ calculus with
first-order Boolean mutable references. %

In \figref{fig:has-type-e}, we recap a canonical binary effect system, distinguishing pure ($\bfalseE$) and impure ($\btrueE$) expressions,
$\btrueE$ taking the role of |@eff| from the examples in the introduction (\secref{sec:challenges}).
The effect ordering is defined as $\subtypeEE{\bfalseE}{\btrueE}$.
We use the operator $\ef\rhd$ for effect composition, which follows the sequential evaluation order of terms.
As $\lambdae$ tracks flow-insensitive store effects,
$\ef{e_1 \rhd e_2}$ is defined as $\ef{e_1 \vee e_2}$ (as shown in \figref{fig:has-type-e}), which
computes the least upper bound of the effects $\ef{e_1}$ and $\ef{e_2}$.

\subsection{Formalization}

The effect typing rules are shown in \figref{fig:has-type-e}.
The typing judgment has the form $\hasTypeE{\GE}{t}{\TE}{e}$, where $\textcolor{red}{e}$ is a binary effect.
It means that under the assumption $\GE$, the term $t$ has type $\TE$ and its evaluation may induce the observable effect $\EF{e}$. 
A function type carries an effect annotation $\textcolor{red}{e}$ for its latent effect,
which is the effect performed when a function of that type is applied.

Rules for mutable references (\textsc{T-Ref-E}, \textsc{T-Get-E} and \textsc{T-Put-E}) are impure expressions, 
each assigned the effect $\btrueE$.
Constant terms (\textsc{T-Cst-E}), variables (\textsc{T-Var-E}) and function abstractions (\textsc{T-Abs-E}) 
are pure, each assigned the effect $\bfalseE$.
For compound terms (\ie, \textsc{T-App-E} and \textsc{T-Bin-E}), the final effect is computed by 
composing the effects of their sub-terms.

\figref{fig:sub-type-e} shows the subtyping rules for $\lambdae$, which are standard.
In particular, the rule for function types (\textsc{S-Abs-E}) demands contravariance in the argument type and covariance in the result type.

The term $\tlet x {t_1} {t_2}$ can be encoded in the standard way 
as ordinary $\lambda$-application $\tapp{(\tabs t_2)}{t_1}$.
Its typing rule is derived from the rules \textsc{T-Abs-E} and \textsc{T-App-E}:

{\centering\small
\infrule[T-Let-E]{
 \hasTypeE {\Gamma} {t_1} {\TE_1} {e_1}
   \qquad
    \hasTypeE {\Gamma, x: \TE_1} {t_2} {\TE_2} {e_2}
  }{
    \hasTypeE {\Gamma}{\tlet {x} {t_1} {t_2}} {\TE_2} { e_1 \rhd e_2}
  }
}%

\subsection{Metatheory}
\begin{figure}[t]\small
\begin{mdframed}
    \judgement{Free Variables}{$\BOX{\FV(t)}$}
    \[
      \begin{array}{l@{\quad}l@{\quad}l@{\quad}l@{\quad}l@{\quad}l@{\quad}l@{\quad}l}
         \FV(b) & = &  \emptyset && 
         \FV(x) & = & \{ x\} \\
         \FV(\tabs{t}) & = & \FV(t) \setminus \{ x \} && 
         \FV(\tapp{t_1}{t_2}) & = & \FV(t_1) \cup \FV(t_2) \\
         \FV(\tref{t}) & = & \FV(t) &&
         \FV(\tget{t}) & = & \FV(t) \\
         \FV(\tput{t_1}{t_2}) & = & \FV(t) \cup \FV(t_2) &&
         \FV(\tseqb{t_1}{t_2}) & = & \FV(t_1) \cup \FV(t_2)
      \end{array}    
    \]
\end{mdframed}    
\caption{Computing free variables of a given term (standard).} 
\label{fig:fv}   
\end{figure} %
\begin{figure}[t]\small
    \begin{mdframed}
        \judgement{Context for Contextual Equivalence}{}
         \[\begin{array}{l@{\ \ }c@{\ \ }l@{\qquad\qquad\ }l@{\ \ }c@{\ \ }l}
				{C} & ::= & \square \mid \tapp{C}{t} \mid  \tapp{t}{C} \mid  \tabs{C} \mid \tref{C} \mid\ \tget{C} \mid \tput{C}{t} \mid \tput{t}{C} \mid \tseqb{t_1}{C} \mid \tseqb{C}{t_2} &  &  & \\
			\end{array}\]
            \judgement{Context Typing Rules}{
                \BOX{C : (\Gamma; T\ \ef{e}) \carrow (\Gamma'; T'\ \ef{e'}\ \flt')}
            }
            
            \infax[c-$\square$]{
                \square: (\Gamma; T\ \ef{e}) \carrow (\Gamma; T\ \ef{e}\ \flt)
            }

            \infrule[c-$\lambda$]{
               C: (\Gamma; T\ \ef{e}) \carrow ((\G', x: T_1); T_2\  \ef{e_2}\ \flt') 
            }{
              \tabs{C}:(\Gamma; T\ \ef{e}) \carrow {\Gamma'; (\TFunE{T_1}{T_2}{e_2})\ \bfalseE \ (\flt' \setminus x)}
            }

            \infrule[c-app-1]{
              C: (\Gamma; T\ \ef{e}) \carrow (\Gamma'; (\TFunE{T_1}{T_2}{e})\ \ef{e_f}\ \flt') \andalso
              \hasTypeE{\Gamma'}{t}{T_1}{\ef{e_1}} 
            }{
              \tapp{C}{t}: (\Gamma; T\ \ef{e}) \carrow (\Gamma'; T_2\ \ef{e_f \rhd e_1 \rhd e }\ \flt')
            }

            \infrule[c-app-2]{
             \hasTypeE{\Gamma'}{t}{\TFunE{T_1}{T_2}{e}}{\ef{e_f}} \andalso 
             C: (\Gamma; T\ \ef{e}) \carrow (\Gamma'; T_1\ \ef{e_1}\ \flt_1) 
            }{
              \tapp{t}{C}: (\Gamma; T\ \ef{e}) \carrow (\Gamma'; T_2\ \ef{e_f \rhd e_1 \rhd e}\ \fv{t})
            }

            \infrule[c-ref]{                 
              C: (\Gamma; T\ \ef{e}) \carrow (\Gamma'; \TBoolE\ \ef{e'}\ \flt')
            }{
              \tref{C}:(\Gamma; T\ \ef{e}) \carrow (\Gamma'; \TRefE\ \btrueE \ \flt')   
            }

            \infrule[c-get]{
              C:(\Gamma; T\ \ef{e}) \carrow (\Gamma'; \TRefA\ \ef{e'}\ \flt')
            }{
                \tget{C}: (\Gamma; T\ \ef{e}) \carrow (\Gamma'; \TBoolE; \btrueE\ \flt')
            }

            \infrule[c-put-1]{
              C: (\Gamma; T\ \ef{e}) \carrow (\Gamma'; \TRefE\ \ef{e_1} \ \flt_1) \andalso 
               \hasTypeE {\Gamma}{t}{\TBoolE}{\ef{e_2}}
            }{
                \tput{C}{t}: (\Gamma; T\  \ef{e}) \carrow (\Gamma'; \TBoolE\ \btrueE\ \flt_1)
            }

            \infrule[c-put-2]{
               \hasTypeE {\Gamma}{t}{\TRefE}{\ef{e_1}} \andalso
               C: (\Gamma; T\ \ef{e}) \carrow (\Gamma'; \TBoolE\ \ef{e_2}\ \flt_2) 
            }{
                \tput{t}{C}: (\Gamma; T\ \ef{e}) \carrow (\Gamma'; \TBoolE\ \btrueE \ \flt_2)
            }

           \infrule[c-bin-1]{
              \hasTypeE{\Gamma}{t}{\TBoolE}{\ef{e_1}} \andalso
              C; (\Gamma; T\ \ef{e}) \carrow (\Gamma'; \TBoolE\ \ef{e_2}\ \flt_2)
            }{
             \tseqb{t}{C}: (\Gamma; T\ \ef{e}) \carrow (\Gamma'; \TBoolE\ \ef{e_1 \rhd e_2} \ \flt_2)
            }

            \infrule[c-bin-2]{
              C; (\Gamma; T\ \ef{e}) \carrow (\Gamma'; \TBoolE\ \ef{e_1}\ \flt_1) \andalso
                \hasTypeE{\Gamma}{t}{\TBoolE}{\ef{e_2}}
              }{
               \tseqb{C}{t}: (\Gamma; T\ \ef{e}) \carrow (\Gamma'; \TBoolE\ \ef{e_1\rhd e_2}\ \flt_1)
              } 
            
            \infrule[c-sub]{
              C: (\Gamma; T\ \ef{e}) \carrow (\Gamma'; T'\ \ef{e'}) \andalso 
              T' <: T'' \andalso 
              \ef{e' <: e''}
            }{
              C_e: (\Gamma; T\ \ef{e}) \carrow (\Gamma'; T''\ \ef{e''})   
            }
    \end{mdframed}
    \caption{Context typing rules for $\lambdae$.}
    \label{fig:context_effs}
\end{figure}
 The core metatheoretic development is fairly standard. 
We define program equivalence in terms of observational equivalence,
and establish it via binary logical relations (details in \secref{sec:lr}).
 
In particular, we say terms $t_1$ and $t_2$ are observationally equivalent, written as 
$\ctxEquivE{\GE}{t_1}{t_2}{\TE}{e}$, if they have the same observable behavior when placed in any program context $C$.
To achieve that, we use the following semantic typing judgment: 
\begin{displaymath}
\small
\semTypeE{\GE}{t_1}{t_2}{\TE}{e} \DEF \forall (\VE,\W) \in \VG{\GE\ \fv{t_1} \cup \fv{t_2}}, (\VE, \W, t_1, t_2) \in \ETE{\TE}{\textcolor{red}{e}},
\end{displaymath}
$\VE$ is a pair of related value environments, $\W$ is a world (defined in~\secref{sec:world}), 
$\VG{\GE\ \fv{t_1} \cup \fv{t_2}}$ is the logical interpretation of typing environments, 
and $\ETE{\TE}{e}$ is the value interpretation of terms, which entails semantic type soundness and termination.
The semantic typing judgment means that terms $t_1$ and $t_2$ are logically related at type $T$ and effects $\EF{e}$ under $\Gamma$.
We define $\fv{t}$ to compute the free variables of term $t$, which is standard, shown in \figref{fig:fv}. 
These definitions are given in terms of their counterparts in $\lambdaae$ (see ~\secref{sec:lr}),
where relevant judgments in $\lambdae$ are encoded into those in $\lambdaae$ (see \secref{sec:encoding}).
The main results are summarized below.

\begin{theorem}[Fundamental Property] If $\hasTypeE{\GE}{t}{\TE}{e}$, 
  then  $\semTypeE{\GE}{t}{t}{\TE}{e}$.
\end{theorem}

\noindent The following theorem shows the soundness of our logical relations \wrt observational equivalence.
\begin{theorem} [Observational Equivalence] 
  if $\hasTypeE{\GE}{t_1}{\TE}{e}$ and $\hasTypeE{\GE}{t_2}{\TE}{e}$, 
  then $\semTypeE{\GE}{t_1}{t_2}{\TE}{e}$ implies $\ctxEquivE{\GE}{t_1}{t_2}{\TE}{e}$.
\end{theorem}

In the following theorem, 
we write $C: (\GE; \TE\ \EF{e}) \carrow (\GEP\ \flt'; \TEP\ \EF{e'})$ to 
mean that the context $C$ is a program, which has type 
$\TEP\ \EF{e'}$ (closed under $\GEP$) and whose free variables are in $\flt'$,  with a hole that can be filled 
with any program of type $\TE\ \EF{e}$ (closed under $\GE$). 
The typing rules for well-typed contexts are shown in \figref{fig:context_effs}.
\begin{theorem}[Effect Safety of Effect Type System] A syntactically pure expression is semantically (observationally) pure:
\[
\small
  \hasTypeE \GE t \TE \bfalse 
  \quad\Rightarrow\quad
  \forall C:(\GE; \TE\ \bfalseE) \carrow (\GEP\ \flt'; \TEP\ \EF{e}).\ \ctxEquivE{\GEP}{\tapp {(\tabs {C[x]})} t}{C[t]}{\TEP}{e}
\]
\end{theorem}

We demonstrate that pure terms have expected algebraic properties in mathematical formulas, \eg, they allow reordering of operand expressions.
\begin{theorem}[Reordering] Pure terms can be evaluated in any order, \ie, swapping arguments of commutative operations is legal:
\[
\small
  \hasTypeE \GE {\tseqb{t_1}{t_2}} \TBoolE \bfalse
  \quad\Rightarrow\quad
  \semTypeE{\GE}{\tseqb{t_1}{t_2}}{\tseqb{t_2}{t_1}}{\TBoolE}{\bfalse}
\]
\end{theorem}

\subsection{Discussion}\label{sec:discuss_effs}

\bfparagraph{Effects are Properties of Evaluating an Expression (Positive)}

Effects classify properties of computations rather than values.
The system is terminating, so we cannot type $\omega$. 
But with a slight extension to allow functions to call themselves recursively, this becomes possible, 
and we can treat recursive calls as effectful. A more refined scheme could detect structurally recursive calls and treat them as pure. This, as one would expect, leads to $\omega$ being syntactically impure and $\tabs \omega$ syntactically pure.
To not complicate the presentation and formalism with multiple kinds of effects, we do not further consider nontermination in this paper. We rather focus on store effects, and the exact same considerations can be made with $\tref \ttrue$ instead of $\omega$.

\bfparagraph{Possibility for Flow-Sensitive Reasoning (Positive)}
We can achieve flow-sensitive reasoning, \eg,
by considering an alternative system with a single global boolean state and two operations: 
\texttt{tic} and \texttt{toc}.
Each operation expects the state value to be set one way and change it to the other, 
or fail if it has the wrong value. 
Thus, legal sequences are \texttt{tic},\texttt{toc},\texttt{tic},... etc.

Such operational behavior can be modeled with a more specialized effect composition operation $e \rhd e'$, 
which captures the legal sequences. 
That is, both $\texttt{tic} \rhd \texttt{toc}$ and $\texttt{toc} \rhd \texttt{tic}$ are well-defined, 
but invalid compositions, \eg, $\texttt{tic} \rhd \texttt{tic}$, are undefined.

\bfparagraph{Allocations are Impure: Hence, they have Effects (Neutral, light negative)}

We discussed earlier that allocations are impure semantically, as they interact with the store, 
and yield nondeterministic store locations.
In our binary effect system, impurity is expressed via the effect annotation $\btrue$.
Thus, allocations are treated the same as read and write effects.

We could refine the effect hierarchy with a more fine-grained hierarchy, 
\ie, distinguishing read, write and allocation effects.
Fundamentally, the refinement still treats allocation as an effect, \ie, as a property of evaluating an expression.
However, from a semantic perspective, what we often want to characterize is not merely that an allocation occurred, 
but rather that \emph{the resulting value is a fresh  location}, 
\ie, a property of the result value \wrt the context, instead.
We will come back to this later.

\bfparagraph{No Effect Masking (Negative)}

There are many terms that are semantically pure but cannot be assigned a syntactically pure type. 
By necessity, any approximation of nontermination cannot be precise. 
And we also do not obtain effect masking: since every allocation has to be effectful, 
an expression such as $\tlet x {\tref t} {\tget x}$ cannot be assigned a pure type syntactically.

Of course, if we are willing to enrich the effect lattice with extra information, 
it becomes possible to model effect masking and other features. 
Benton et al.'s work~\cite{DBLP:conf/popl/Benton0N14}, \eg, achieves this by considering effects relative to store regions, and masking out effects on regions that are no longer observable, \ie, are no longer referenced anywhere in the term or the environment.

\bfparagraph{No Effect Polymorphism (Negative)}

Our simple effect system shows its limitations in the presence of 
higher-order functions (or objects with dynamic method dispatch). 
As in \secref{sec:challenges}, consider a function such as |map(f)(ls)|, 
whose effect depends on the effect of the function argument |f|. 
We can type |map| with the following signature ($T_1$ and $T_2$ are
fixed but arbitrary types):
\[\small
\setlength{\abovedisplayskip}{0pt}
\setlength{\belowdisplayskip}{0pt}
\setlength{\abovedisplayshortskip}{0pt}
\setlength{\belowdisplayshortskip}{0pt}
\begin{array}{rcl}
map & : & (\TFunE {T_1} {T_2} {\btrueE})\ \stackrel{\textcolor{red}{\bfalseE}}{\to} 
          (\TFunE {\TListAE{T_1}} {\TListAE{T_2}} {\btrueE}) \\
\end{array}
\]
Clearly, our binary effect system lacks the expressiveness to abstract over effects.
While |map(f)| always has pure effect $\bfalseE$, |map(f)(ls)| always has type $\TListAE{T_2} {\btrueE}$ even when $f$ is pure.
A natural extension would be to adopt a polymorphic effect system~\cite{DBLP:conf/popl/LucassenG88, DBLP:journals/iandc/TalpinJ94},
however, as discussed in \secref{sec:challenges}, this would come at a much greater expense in complexity for users. 
In essence, every higher-order function needs additional effect parameters to abstract over the effect of the function arguments, and ultimately,
a function's effect must include all the effects of all functions it may transitively invoke~\cite{DBLP:phd/ch/Rytz14,DBLP:journals/toplas/BoruchGruszeckiOLLB23}.

\bfparagraph{Conclusion}
In terms of expressiveness, we note that there are relevant classes of 
semantically pure terms that $\lambdae$ cannot type as syntactically 
pure, specifically most examples of effect polymorphism such as |map| and effect
masking such as $\tlet x {\tref t} {\tget x}$. 
On the plus side, however, effects support a clear use vs.\ mention
distinction ($\tlet x y x$ where $y: \TRef \in \Gamma$ is pure), and effects 
can naturally model flow-sensitive behavior through sequential composition~\cite{DBLP:journals/toplas/Gordon21}, \eg, ownership transfer and deallocation~\cite{DBLP:journals/corr/abs-2510-08939}.
As we will show in the following \secref{sec:ability}, 
ability systems behave in the opposite way on these examples.

\section{Ability Type Systems ($\lambdaa$)}\label{sec:ability}

As an alternative to effect type systems, systems that track the flow of capabilities have received significant interest in recent years. 
Rather than treating effects as a property of how a term evaluates, these systems track the flow of \emph{capabilities} or \emph{resources} (such as mutable references).

This, most importantly, has the promise of dealing more gracefully with effect polymorphism: capabilities can be passed down the callgraph through the environment without polluting the types of higher-order functions with additional quantifiers.

A number of concrete type systems with these goals in mind have been proposed, 
\eg, the work of~\citet{DBLP:conf/ecoop/XhebrajB0R22} and~\citet{osvald2016gentrification}, Scala Capturing Types~\cite{DBLP:journals/toplas/BoruchGruszeckiOLLB23}, 
and Reachability Types~\cite{DBLP:journals/pacmpl/BaoWBJHR21,DBLP:journals/pacmpl/WeiBJBR24}. 
However, a fundamental study of their effect semantics has not been conducted.
In the following, we present and analyze a canonical binary ($\AB{a} := \bfalseA \mid \btrueA$) ``ability'' type system from our perspective of observational purity.

\subsection{Formalization}\label{sec:ability_formalization}

\begin{figure}[t]\small
    \begin{mdframed}[leftmargin=2pt, rightmargin=2pt]
    \typicallabel{T-False-A}
    \judgement{Types \& Ability}{$\BOX{\TA::=...}\ \BOX{\textcolor{blue}{a}::=...}$} \\\vspace{-3ex}
    \[
      \begin{array}{l@{\quad}l@{\quad}l@{\quad}l@{\quad}l@{\quad}l@{\quad}l@{\quad}l@{\quad}l@{\quad}l}
        \TA & := & \TBoolA \mid \TRefA \mid \TFunA {\TA_1} {a_1} {\TA_2} {a_2} & & & \textcolor{blue}{a} & := & \bfalseA \mid \btrueA   
      \end{array}%
    \]%
    \vspace{-4ex}\\
    \judgement{Term Typing}{$\BOX{\hasTypeA \GA t \TA a}$}\vspace{-4ex}
    
    \begin{multicols}{2}
    
    \infax[T-Cst-A]{
      \hasTypeA \GA {b} \TBoolA \bfalseA
    }
    
    \infrule[T-Bin-A]{
      \hasTypeA \GA {t_1} \TBoolA {a_1} \\
      \hasTypeA \GA {t_2} \TBoolA {a_2}
    }{
      \hasTypeA {\GA} {\tseqb{t_1}{t_2}} {\TBoolA}{\bfalseA}
    }
    
    \infrule[T-Ref-A]{
      \hasTypeA \GA t \TBoolA a
    }{
      \hasTypeA \GA {\tref t} \TRefA \btrueA
    }
    
    \infrule[T-Get-A]{
      \hasTypeA \GA t \TRefA a
    }{
      \hasTypeA \GA {\tget t} \TBoolA \bfalseA
    }
    
    \infrule[T-Put-A]{
      \hasTypeA \GA {t_1} \TRefA {a_1} \\
      \hasTypeA \GA {t_2} \TBoolA {a_2}
    }{
      \hasTypeA \GA {\tput {t_1} {t_2}} \TBoolA \bfalseA
    }
    
    \columnbreak
    
    \infrule[T-Var-A]{
      \indexrTypeA \GA x \TA \AB{a}
    }{
      \hasTypeA \GA {\tvar x} \TA a
    }
    
    \infrule[T-Abs-A]{
      \hasTypeA {\GA, x : \TA_1\ \AB{a_1}} t {\TA_2} {\AB{a_2}} \\
      \envCapA \GA {\fv t \setminus x} {a_f}
    }{
      \hasTypeA \GA {\tabs t} {(\TFunA {\TA_1} {a_1} {\TA_2} {a_2})} {a_f}
    }
    
    \infrule[T-App-A]{
      \hasTypeA \GA {t_1} {(\TFunA {\TA_1} {a_1} {\TA_2} {a_2})} {a_f} \\
      \hasTypeA \GA {t_2} {\TA_1} {a_1}
    }{
      \hasTypeA \GA {\tapp {t_1} {t_2}} {\TA_2} {a_2}
    }
    
    \infrule[T-Sub-A]{
      \hasTypeA \GA t {\TA_1} {a_1} \andalso
      \subtypeTA{\TA_1}{\TA_2} \\ 
      \subtypeAA{a_1}{a_2}
    }{
      \hasTypeA \GA t {\TA_2} {a_2}
    }
    
    \end{multicols}
    \vspace{-2ex}
    \judgement{Augmented Term Typing with Ambient Ability}{$\BOX{\hasTypePhiA \GA {a'} t \TA a}$}\vspace{-3ex}
    
    \infrule{
      \hasTypeA \GA t \TA a \andalso
      \envCapA \GA {\fv t} {a'}
    }{
      \hasTypePhiA \GA {a'} t \TA a
    }

    \end{mdframed}

    \caption{Ability Type System $\lambda_a$.}
    \label{fig:has-type-a}
    \end{figure} %
\begin{figure}[t]\small
    \begin{mdframed}
    \typicallabel{T-False-A}
    \judgement{Subtyping}{$\BOX{\subtypeTA{\TA_1}{\TA_2}}\ \BOX{\subtypeAA{a_1}{a_2}}$}\vspace{-2ex}
    \begin{multicols}{2}
       \infax[S-Cst-A]{
         \subtypeTA{\TBoolA}{\TBoolA}
       }

       \infax[S-Tref-A]{ 
        \subtypeTA{\TRefA}{\TRefA}
       }
       
   \end{multicols}

   \infrule[S-Abs-A]{
        \subtypeTA{\TA_3}{\TA_1} \andalso
        \subtypeTA{\TA_2}{\TA_4} \andalso
        \subtypeAA{a_3}{a_1} \andalso
        \subtypeAA{a_2}{a_4} 
     }{
        \subtypeTA{(\TFunA{\TA_1}{a_1}{\TA_2}{a_2})}{(\TFunA{\TA_3}{a_3}{\TA_4}{a_4})}
     }

     \begin{multicols}{3}
      \infax[S-A]{
        \subtypeAA{\bfalse}{\btrue}
       }

     \infax[S-$\bfalse$-A]{
      \subtypeAA{\bfalse}{\bfalse}
     }

     \infax[S-$\btrue$-A]{
      \subtypeAA{\btrue}{\btrue}
     }

     \end{multicols}
    \end{mdframed}
    \caption{Ability Subtyping Rules $\lambda_{a}$.} 
    \label{fig:sub-type-a}
\end{figure} 
The ability typing rules are shown in \figref{fig:has-type-a}.
The ability typing judgment has the form $\hasTypeA \GA t \TA a$, where $\AB{a}$ is a binary ability annotation,
with $\btrueA$ taking the role of |@cap| from \secref{sec:challenges}.

\bfparagraph{Ability Qualifier}
The \emph{ability qualifier} $\ab{a}$ in a judgment $\hasTypeA \GA t \TA a$ describes 
if evaluating $t$ yields a \emph{resource} -- 
that is, either a reference or a value that may transitively reach other resources. 
The typing rules ensure that resources are typed with ability qualifier $\btrueA$ wherever they flow.

\bfparagraph{Ambient Ability} 
The \emph{ambient ability} $\ab{a}$ of a term $t$ is based on the set of resources from the environment needed to evaluate $t$. Specifically, it is an upper bound on the ability qualifiers of $t$'s free variables:
$
  \envCapA \GA {\fv t} {a}
$.
In the typing rules, the key to tie these two notions together is in rule \textsc{(T-Abs-A)}: 
the hypothesis $\envCapA \GA {\fv t \setminus x} {a_f}$  computes the ability qualifer $a_f$ assigned to the function from the ambient ability of the body expression, \ie, the function becomes itself a resource if it captures any resources (ability qualifier $\ne \bfalseA$) from the environment. 

For economy of notation, we often use an augmented term typing judgment that includes the ambient ability $a'$ in the form of
$\hasTypePhiA \GA {a'} t \TA a$.

\figref{fig:sub-type-a} shows the subtyping rules for $\lambdaa$, which are standard.

The term $\tlet x {t_1} {t_2}$ can be encoded in the standard way 
as ordinary $\lambda$-application $\tapp{(\tabs t_2)}{t_1}$.
Its typing rule is derived from the rules \textsc{T-Abs-A} and \textsc{T-App-A}:

{\centering\small
\infrule[T-Let-A]{
 \hasTypeA {\Gamma} {t_1} {T_1} {a_1}
   \qquad
    \hasTypeA {\Gamma, \TA_1\ \AB{a_1}} {t_2} {\TA_2} {a_2}
  }{
    \hasTypeA {\Gamma}{\tlet {x} {t_1} {t_2}} {\TA_2} {a_2}
  }
}%

\bfparagraph{Syntactic Purity}

Which expressions are syntactically pure? Returning a resource is not per se impure, but returning a freshly allocated one is. %
Since there is no distinction in the types, we have to treat any expression with type $\TA\ \btrueA$ as impure.
To determine potential uses of resources as part of evaluation, we have to consider the ambient ability through the free variables of an expression. An expression $t$ with ambient ability $\btrueA$, 
\ie, $\ab{\envCapTrueA \GA {\fv t}}$, is also impure. 
Thus, pure expressions are characterized by the augmented typing $\hasTypePhiA \GA \bfalseA t \TA \bfalseA$.

\subsection{Metatheory}
\begin{figure}[t]\small
    \begin{mdframed}
        \judgement{Context for Contextual Equivalence}{}
         \[\begin{array}{l@{\ \ }c@{\ \ }l@{\qquad\qquad\ }l@{\ \ }c@{\ \ }l}
				{C} & ::= & \square \mid \tapp{C}{t} \mid  \tapp{t}{C} \mid  \tabs{C} \mid \tref{C} \mid\ \tget{C} \mid \tput{C}{t} \mid \tput{t}{C} \mid \tseqb{t_1}{C} \mid \tseqb{C}{t_2} &  &  & \\
			\end{array}\]
            \judgement{Context Typing Rules}{
                \BOX{C : (\Gamma\ \ab{a_\flt}; T\ \ab{a}) \carrow (\Gamma'\ \flt'; T'\ \ab{a'})}
            }
            
            \infax[c-$\square$]{
                \square: (\Gamma\ \ab{a_\flt}; T\ \ab{a}) \carrow (\Gamma\ \flt; T\ \ab{a})
            }

            \infrule[c-$\lambda$]{
               C: (\Gamma\ \ab{a_\flt}; T\ \ab{a}) \carrow ((\G', x: T_1\ \ab{a_1})\ \flt'; T_2\ \ab{a_2}) \andalso
               \envCapA{\Gamma'}{\flt' \setminus x}{a_f'} 
            }{
              \tabs{C}:(\Gamma\ \ab{a_\flt}; T\ \ab{a}) \carrow {\Gamma'\ (\flt' \setminus x); (\TFunA{T_1}{a_1}{T_2}{a_2})\ \ab{a_f'\vee a_{\flt}}}
            }

            \infrule[c-app-1]{
              C: (\Gamma\ \ab{a_\flt}; T\ \ab{a}) \carrow (\Gamma'\ \flt'; (\TFunA{T_1}{a_1}{T_2}{a_2})\ \ab{a_f}) \andalso
              \hasTypeA{\Gamma'}{t}{T_1}{\ab{a_1}}
            }{
              \tapp{C}{t}: (\Gamma\ \ab{a_\flt}; T\ \ab{a}) \carrow (\Gamma'\ \flt'; T_2\ \ab{a_f \vee a_1 \vee a_2})
            }

            \infrule[c-app-2]{
             \hasTypeA{\Gamma'}{t}{\TFunA{T_1}{a_1}{T_2}{a_2}}{\ab{a_f}} \andalso 
             C: (\Gamma\ \ab{a_\flt}; T\ \ab{a}) \carrow (\Gamma'\ \flt_1; T_1\ \ab{a_1}) 
            }{
              \tapp{t}{C}: (\Gamma\ \ab{a_\flt}; T\ \ab{a}) \carrow (\Gamma'\ \fv{t}; T_2\ \ab{a_f \vee a_1 \vee a_2})
            }

            \infrule[c-ref]{                 
               C: (\Gamma\ \ab{a_\flt}; T\ \ab{a}) \carrow (\Gamma'\ \flt'; \TBoolA\ \ab{a'})
            }{
              \tref{C}:(\Gamma\ \ab{a_\flt}; T\ \ab{a}) \carrow (\Gamma' \ \flt'; \TRefA\ \btrueA)   
            }

            \infrule[c-get]{
                C:(\Gamma\ \ab{a_\flt}; T\ \ab{a}) \carrow (\Gamma'\ \flt'; \TRefA\ \ab{a'})
            }{
                \tget{C}: (\Gamma\ \ab{a_\flt}; T\ \ab{a}) \carrow (\Gamma' \ \flt'; \TBoolA; \bfalseA)
            }

            \infrule[c-put-1]{
               C: (\Gamma\ \ab{a_\flt}; T\ \ab{a}) \carrow (\Gamma'\ \flt_1; \TRefA\ \ab{a_1}) \andalso 
               \hasTypeA {\Gamma}{t}{\TBoolA}{\ab{a_2}}
            }{
                \tput{C}{t}: (\Gamma\ \ab{a_\flt}; T\ \ab{a}) \carrow (\Gamma'\ \flt_1; \TBoolA\ \bfalseA)
            }

            \infrule[c-put-2]{
               \hasTypeA {\Gamma}{t}{\TRefA}{\ab{a_1}} \andalso
               C: (\Gamma\ \ab{a_\flt}; T\ \ab{a}) \carrow (\Gamma'\ \flt_2; \TBoolA\ \ab{a_2}) 
            }{
                \tput{t}{C}: (\Gamma\ \ab{a_\flt}; T\ \ab{a}) \carrow (\Gamma' \  \flt_2; \TBoolA\ \bfalseA)
            }

           \infrule[c-bin-1]{
              \hasTypeA{\Gamma}{t}{\TBoolA}{\ab{a_1}} \andalso
              C; (\Gamma\ \ab{a_\flt}; T\ \ab{a}) \carrow (\Gamma'\ \flt_2; \TBoolA\ \ab{a_2})
            }{
             \tseqb{t}{C}: (\Gamma\ \ab{a_\flt}; T\ \ab{a}) \carrow (\Gamma' \ \flt_2; \TBoolA\ \bfalseA)
            }

            \infrule[c-bin-2]{
                C; (\Gamma\ \ab{a_\flt}; T\ \ab{a}) \carrow (\Gamma'\ \flt_1; \TBoolA\ \ab{a_1}) \andalso
                \hasTypeA{\Gamma}{t}{\TBoolA}{\ab{a_2}}
              }{
               \tseqb{C}{t}: (\Gamma\ \ab{a_\flt}; T\ \ab{a}) \carrow (\Gamma'\ \flt_1; \TBoolA\ \bfalseA)
              }
            
            \infrule[c-sub]{
              C: (\Gamma\ \ab{a_{\flt}}; T\ \ab{a}) \carrow (\Gamma'\ \flt'; T'\ \ab{a'}) \andalso
              T' <: T'' \andalso 
              \ab{a' <: a''}
            }{
              C: (\Gamma\ \ab{a_{\flt}}; T\ \ab{a}) \carrow (\Gamma'\ \flt'; T''\ \ab{a''})
            }

    \end{mdframed}
    \caption{Context typing rules for $\lambdaa$. The definition of context is repeated for easy reference.}
    \label{fig:context_ab}
\end{figure}
 
Similar to $\lambdae$, the core metatheory for $\lambdaa$ is also standard.
We define program equivalence in terms of observational equivalence and establish it via the binary logical relations for $\lambdaa$ (details in \secref{sec:lr}).

In particular, we say terms $t_1$ and $t_2$ are observationally equivalent, written as 
$\ctxEquivA{\GA}{t_1}{t_2}{\TA}{a}$, if they have the same observable behavior when placed in any program context $C$.
To achieve that, we use the following semantic typing judgment:
\begin{displaymath}
\small
\semTypeA{\GA}{t_1}{t_2}{\TA}{a} \DEF \forall (\VE,\W) \in \VG{\GA \ \fv{t_1} \cup \fv{t_2}}, (\VE, \W, t_1, t_2) \in \ETA{\TA}{a},
\end{displaymath}
$\VE$ is a pair of related value environments, $\W$ is a world (defined in~\secref{sec:world}), 
$\VG{\GA \ \fv{t_1} \cup \fv{t_2}}$ is the logical interpretation of typing environment, 
and $\ETA{\TA}{a}$ is the value interpretation of terms, which entails semantic  soundness and termination.
The semantic typing judgment means that terms $t_1$ and $t_2$ are logically related at type T and ability $\ab{a}$ under $\Gamma$.
These definitions are given in terms of their counterparts in $\lambdaae$ (~\secref{sec:lr}),
where relevant judgments in $\lambdaa$ are encoded into those in $\lambdaae$ (see \secref{sec:encoding}).
The main results are summarized below.

\begin{theorem}[Fundamental Property] If $\hasTypeA{\GA}{t}{\TA}{a}$, 
  then  $\semTypeA{\GA}{t}{t}{\TA}{a}$.
\end{theorem}

\noindent The following theorem shows the soundness of our logical relations \wrt observational equivalence.
\begin{theorem} [Observational Equivalence]
  if $\hasTypeA{\GA\ \ab{a_{\flt}}}{t_1}{\TA}{a}$ and $\hasTypeA{\GA\ \ab{a_{\flt}}}{t_2}{\TA}{a}$, 
  then $\semTypeA{\GA}{t_1}{t_2}{\TA}{a}$ implies $\ctxEquivA{\GA\ \ab{a_{\flt}}}{t_1}{t_2}{\TA}{a}$.
\end{theorem} 

In the following theorem, 
we write $C: (\GA\ \ab{a_{\flt}}; \TA\ \AB{a}) \carrow (\GAP\ \flt'; \TAP\ \AB{a'})$ to 
mean that the context $C$ is a program, 
which has type $\TAP\ \AB{a'}$ (closed under $\GAP$) and whose free variables are in $\flt'$,  with a hole that can be filled 
with any program $t$, such that $\GA\ \ab{a_{\flt}} \vdash_{\rm{A}}~ t: \TA\ \AB{a}$. 
\figref{fig:context_ab} shows the context typing rules for $\lambdaa$.

\begin{theorem}[Effect Safety of Ability Type System] A syntactically pure expression is semantically (observationally) pure:
\[
\small
  \hasTypePhiA \GA \bfalseA t \TA \bfalseA 
  \quad\Rightarrow\quad
  \forall C:(\GA\ \bfalseA; \TA\ \bfalseA) \carrow (\GAP\ \flt'; \TAP\ \AB{a}).\ \ctxEquivA{\GAP}{\tapp {(\tabs {C[x]})} t}{C[t]}{\TAP}{a}
\]
\end{theorem}

We demonstrate that pure terms have expected algebraic properties in mathematical formulas, \eg, they allow reordering of operand expressions.

\begin{theorem}[Reordering] Pure terms can be evaluated in any order, \ie, swapping arguments of commutative operations is legal:
\[
\small
  \hasTypePhiA \GA \bfalseA {\tseqb{t_1}{t_2}} \TBoolA \bfalseA
  \quad\Rightarrow\quad
  \semTypeA{\GA \bfalseA}{\tseqb{t_1}{t_2}}{\tseqb{t_2}{t_1}}{\TBoolA}{\bfalseA}
\]
\end{theorem}

\subsection{Discussion}\label{sec:diss_ability}

\bfparagraph{No Use/Mention Distinction (Negative)}

We track the flow of resources, but not precisely when they are used. So typing-wise (in a suitable extension with recursive functions), the terms $\omega$ and $\tabs \omega$ are both resources and hence syntactically impure. 
This may seem like a weakness compared to effect systems, but it is also what enables seamless effect polymorphism.

\bfparagraph{Effect Polymorphism I (Positive)}

A notable advantage of capability-based systems is that 
they provide effect polymorphism without requiring explicit effect quantification. 
Consider again the function |map(f)(ls)| discussed earlier (in Sections~\ref{sec:challenges} and~\ref{sec:discuss_effs}). In $\lambdaa$ we can type |map| as:
\[\small
\begin{array}{rcl}
map & : & \TFunA {(\TFunA {T_1} {\bfalseA} {T_2} {\bfalseA})} {\btrueA} %
          {(\TFunA {\TListAE{T_1}} {\bfalseA} {\TListAE{T_2}}  {\bfalseA})} {\btrueA} \\
\end{array}
\]
Now, the partial application |map(f)| has type $(... \to ...)\ \btrueA$ and is thus
impure (no use/mention distinction!). However, rule \textsc{T-App-A} ignores the ability of the function and thus
|map(f)(ls)| has type $\TListAE{T_2}\ \bfalseA$. When is |map(f)(ls)| pure? 
Precisely when |map(f)(ls)| has \emph{ambient ability} $\bfalseA$,
\ie, when neither |f| nor |ls| mentions any capabilities from the context.
As a result, capability systems achieve a lightweight form of effect polymorphism
while avoiding the complexity %
associated with polymorphic effect systems.

\bfparagraph{Effect Polymorphism II (Neutral)}

The preceding discussion concerns passing resources \emph{down} the call chain.
In contrast, passing resources \emph{up} the call chain presents different challenges.
As mentioned above, we cannot treat a function that may return a resource as pure (because it might return a fresh resource).
If we want to return any capability up the chain (even pre-existing ones), the expression must be treated as impure.

\bfparagraph{Effect Masking (Positive)}
The ability system $\lambdaa$ masks effects that cannot be observed outside of a given expression.
For example, the expression $\tlet x {\tref t} {\tget x}$ is syntactically pure in $\lambdaa$,
as it does not return a resource, and its ambient ability is $\bfalseA$.

\bfparagraph{Beyond Boolean Refs (Positive)}

We can generalize mutable references from $\TBoolA$ to other non-resource types (\eg, values of closed $\lambda$-terms),  
and still maintain termination and other properties of the system. 
This would be harder to achieve in an effect system.

\bfparagraph{Conclusion}
In terms of expressiveness, we observe that $\lambdaa$ has opposing strengths
compared to $\lambdae$ (\secref{sec:effects}). Examples of effect polymorphism
such as |map| and effect masking such $\tlet x {\tref t} {\tget x}$ were
impure in $\lambdae$ but are pure in $\lambdaa$. On the flip side, however,
pure ability systems such as $\lambdaa$ do not support a
use vs.\ mention distinction ($\tlet x y x$ where $y: \TRef\ \btrueA \in \Gamma$ is 
impure, but was pure in $\lambdae$) and do not readily model any
sequential or flow-sensitive behavior of programs. 
With the above examples, we have shown that $\lambdaa$ and 
$\lambdae$ are incomparable in terms of expressiveness: each system
has pure terms that are impure in the other, so neither
system subsumes the other. %

\section{Type, Ability, and Effect Systems ($\lambda_{ae}$)}
\label{sec:tae}

We present $\lambda_{ae}$, which integrates ability and effects in a unified system. 
Then we show that $\lambda_{ae}$ can encode $\lambda_e$ and $\lambda_a$.
The meta-theory of $\lambda_{ae}$ will be presented in \secref{sec:lr}. %

\subsection{Formalization}
\begin{figure}[t]\small
    \begin{mdframed}[leftmargin=2pt, rightmargin=2pt]
    \typicallabel{T-False-A}
    \judgement{Types, Ability and Effects}{$\BOX{T::=...}\ \BOX{\textcolor{blue}{a}::=...}\ \BOX{\textcolor{red}{e}::=...}$} \\\vspace{-2ex}
    \[
      \begin{array}{l@{\quad\quad\quad\quad\quad\quad}l@{\quad}l@{\quad}l@{\quad}l@{\quad}l@{\quad}l@{\quad}l@{\quad}l@{\quad}l@{\quad}l@{\quad}l@{\quad}l@{\quad}l@{\quad}l@{\quad}l@{\quad}l}
        & T & := & \multicolumn{6}{l}{\!\!\!\TBoolA \mid \TRefA \mid \TFunAE {T_1} {a_1} {T_2} {a_2}{e}} \\
        & \textcolor{blue}{a} & := & \avc{f}{s} & & 
        \textcolor{blue}{f},\textcolor{blue}{s} & := & \bfalseA \mid \btrueA & &
        \avbot & := & \av \bfalseA \bfalseA &&
        \textcolor{blue}{\bar{a}} & := & \textcolor{blue}{a.f \vee a.s} \\
        & \textcolor{red}{e} & := & \bfalseE \mid \btrueE &&
        \textcolor{red}{e \rhd e'} & := & \textcolor{red}{e \vee e'} &&
        \avtop & := & \av \btrueA \btrueA \\
      \end{array}%
    \]%
    \vspace{-4ex}\\
    \judgement{Term Typing}{$\BOX{\hasTypeAEC \Gamma t T {a} e}$}\vspace{-4ex}
    
    \begin{multicols}{2}
    
    \infax[T-Cst-AE]{
      \hasTypeAEC\Gamma {b} \TBoolA \avbot \bfalseE
    }
    
    \infrule[T-Bin-AE]{
      \hasTypeAEC \Gamma {t_1} \TBoolA {a_1} {e_1} \\
      \hasTypeAEC \Gamma {t_2} \TBoolA {a_2} {e_2} 
    }{
      \hasTypeAEC {\Gamma} {\tseqb{t_1}{t_2}} {\TBoolA} {\avbot} {e_1 \rhd e_2}
    }
    
    \infrule[T-Ref-AE]{
      \hasTypeAEC \Gamma t \TBoolA {a} {e}
    }{
      \hasTypeAEC \Gamma {\tref t} \TRefA {\av \btrue \bfalse} {e}
    }
    
    \infrule[T-Get-AE]{
      \hasTypeAEC \Gamma t \TRefA {a} {e}
    }{
      \hasTypeAEC \Gamma {\tget t} \TBoolA \avbot {e \rhd a.s}
    }

    \infrule[T-Put-AE]{
      \hasTypeAEC \Gamma {t_1} \TRefA {a_1} {e_1}\\
      \hasTypeAEC \Gamma {t_2} \TBoolA {a_2} {e_2}
    }{
      \hasTypeAEC \Gamma {\tput {t_1} {t_2}} \TBoolA \avbot {e_1 \rhd e_2 \rhd {a_1.s}}
    }
    
    \columnbreak
    
    \infrule[T-Var-AE]{
      \indexrTypeA \Gamma x T \textcolor{blue}{a}
    }{
      \hasTypeAEC \Gamma {\tvar x} T {\avc \bfalse {\bar a}} \bfalse
    }
    
    \infrule[T-Abs-AE]{
      \hasTypeAEC {\Gamma, x : T_1\ \textcolor{blue}{a_1}} t {T_2} {a_2} {e_2} \\
      \envCapA \Gamma {\fv t \setminus x} {a_f} \\
      \textcolor{blue}{a_f' = }\ \avc {\bfalse}{{{\bar a}_f}\land (a_2.s \vee e_2)}
    }{
      \hasTypeAEC \Gamma {\tabs t} {(\TFunAE {T_1} {a_1} {T_2} {a_2} {e_2})} {a_f'} {\bfalse}
    }

    \infrule[T-Sub-AE]{
      \hasTypeAEC \Gamma t {T_1} {a_1} {e_1} \andalso
      \subtypeTA{T_1}{T_2} \\ 
      \subtypeAA{a_1}{a_2} \andalso
      \subtypeEE{e_1}{e_2}
    }{
      \hasTypeAEC \Gamma t {T_2} {a_2} {e_2}
    }
    
    \end{multicols}

    \infrule[T-App-AE]{
      \hasTypeAEC \Gamma {t_1} {(\TFunAE {T_1} {a_1} {T_2} {a_2} {e_2})} {a_f} {e_f} \andalso
      \hasTypeAEC \Gamma {t_2} {T_1} {a_1} {e_1} 
    }{
      \hasTypeAE \Gamma {\tapp {t_1} {t_2}} {T_2} 
        {\avc 
          {a_2.f \vee a_2.s \wedge (a_f.f \vee a_1.f)}
          {a_2.s \wedge (a_f.s \vee a_1.s)}}
        {\\ \ef{e_f\ \rhd\ e_1\ \rhd\ (e_2 \wedge (a_f.s \vee a_1.s))} }
    }

     \end{mdframed}
    \caption{Ability and Effect Type System $\lambda_{ae}$. 
    We write $a.f$, and $a.s$ to refer to the fresh and stored components in ability qualifier $a$, respectively. 
    }
    \label{fig:has-type-ae}
    \end{figure} %
\begin{figure}[t]\small
    \begin{mdframed}
    \typicallabel{T-False-A}
    \judgement{Subtyping}{$\BOX{\subtypeTA{T}{T'}}\ \BOX{\subtypeAA{a}{a'}}$}\vspace{-2ex}
    \begin{multicols}{2}
       \infax[S-Cst-AE]{
         \subtypeTA{\TBoolA}{\TBoolA}
       }
       \columnbreak
      \infax[S-Tref-AE]{ 
       \subtypeTA{\TRefA}{\TRefA}
      }
    \end{multicols}%
    \infrule[S-Abs-AE]{
        \subtypeTA{T_3}{T_1} \andalso
        \subtypeTA{T_2}{T_4} \andalso
        \subtypeAA{a_3}{a_1} \andalso
        \subtypeAA{a_2}{a_4} \andalso
        \subtypeEE{e_1}{e_2}
     }{
        \subtypeAA{(\TFunAE{T_1}{a_1}{T_2}{a_2}{e_1})}{(\TFunAE{T_3}{a_3}{T_4}{a_4}{e_2})}
     }%

     \begin{multicols}{3}
        \infax[S-E-AE] {
          \subtypeEE{\bfalse}{\btrue}
        }
    
        \infax[S-E$\bfalse$-AE]{
          \subtypeEE{\bfalse}{\bfalse}
        }
    
        \infax[S-E$\btrue$-AE]{
          \subtypeEE{\btrue}{\btrue}
        }
    
     \end{multicols}%

    \begin{multicols}{3}
      \infax[S-A-AE]{
       \subtypeTA{\bfalseA}{\btrueA}
      }
      
    \infax[S-A$\bfalse$-AE]{
     \subtypeAA{\bfalseA}{\bfalseA}
    }

    \infax[S-A$\btrue$-AE]{
     \subtypeAA{\btrueA}{\btrueA}
    }
  \end{multicols}
  \infrule[S-Sub-AE]{
     \subtypeAA{f_1}{f_2} \andalso 
     \subtypeAA{a_1}{a_2} \
    }{
     \subtypeAA{\avc{f_1}{a_1}} {\av{f_2}{a_2}}
    }
  
    \end{mdframed}
    \caption{Ability and Effect Subtyping System $\lambda_{ae}$.} 
    \label{fig:sub-type-ae}
\end{figure} 
The typing rules are shown in \figref{fig:has-type-ae}.
The typing judgment is in the form of $\hasTypeAEC \Gamma t T a e$, where \textcolor{blue}{$a$} is 
an ability and \textcolor{red}{$e$} is an effect. 
Contrary to $\lambda_a$ presented in \secref{sec:ability}, ability qualifiers $\textcolor{blue}{a}$ now are in the
form of $\textcolor{blue}{\av f s}$, denoting whether the underlying characterized locations may be
 \emph{fresh} or \emph{stored}. %
We often write $\ab{a.f}$ and $\ab{a.s}$ to refer to its fresh and stored components, respectively.%
\footnote{One could add freshness and effects to $\lambdaa$ in two separate steps; we omit the intermediate step for space reasons.}

The fresh component indicates whether an evaluation result may reach new locations allocated during execution.
If the value may reach those locations, they must be tracked as they may be accessed by the evaluation context;
in this case, the \emph{fresh} component is assigned $\btrueA$. 
If no allocation occurs during evaluation, or 
any freshly allocated locations are not reachable from the result (and thus unobservable), 
the component is assigned $\bfalseA$, and these locations can be safely ignored by the reasoning.
See formulas $\recdx{F1}$ and $\recdx{F2}$ in the definition of binary logical relations in \secref{sec:lr_def}.

The stored component indicates whether the result of an evaluation may reach locations that existed prior to the execution. 
In this case, effects on those locations are externally observable, and thus must be tracked as effect qualifier.   
This mechanism is exemplified in rules \textsc{(T-Get-AE)} and \textsc{(T-Put-AE)}, 
where effects on fresh cells are ignored, while effects on stored cells are tracked.
The rule \textsc{(T-Var-AE)} assigns ability qualifier ${\textcolor{blue}{\av \bfalse {(f \vee s)}}}$,
since variables references are by definition to a pre-existing value, thus they cannot be fresh.
Freshness is reserved for allocations during evaluation of a term, but during
evaluation of a variable reference no allocation happens.

The function application rule \textsc{(T-App-AE)} ensures that if a function has
a latent effect $\textcolor{red}{e_2}$, then either the function itself or the argument carries capabilities. 
Those are the only values the function body has access to, thus any
locations it may access %
must be reachable through those values.

The function abstraction rule \textsc{(T-Abs-AE)} specifies that a function is treated as a resource only if 
its value carries either stored or fresh capabilities, and either has latent effects or returns a value that carries capabilities from the environment.

\figref{fig:sub-type-ae} shows the subtyping rules for $\lambdaae$. 
The rule \textsc{s-sub-ae} defines the general sub-typing relation for ability qualifiers, 
where subtyping is applied componentwise to each of the qualifier's components.
For each component, the subtyping relations are defined by the rules \textsc{s-a-ae}, \textsc{s-a$\bfalse$-ae} and \textsc{s-a$\btrue$-ae}.
The rule for function types (\textsc{S-Abs-AE}) demand contravariance in the argument type and the effects, and covariance in the result type.

The term $\tlet x {t_1} {t_2}$ can be encoded in the standard way 
as ordinary $\lambda$-application $\tapp{(\tabs t_2)}{t_1}$.
Its typing rule is derived from the rules \textsc{T-Abs-AE} and \textsc{T-App-AE}.

{\centering\small
\infrule[T-Let-AE]{
 \hasTypeAEC {\Gamma} {t_1} {T_1} {a_1} {e_1}
   \qquad
    \hasTypeAEC {\Gamma, x:T_1} {t_2} {T_2} {a_2} {e_2}
    \qquad 
    \envCapA {\Gamma} {\fv {t_2} \setminus x} {a_f} 
  }{
    \hasTypeAE {\Gamma}{\tlet {x} {t_1} {t_2}} {T_2} {\avc {(a_1.f \wedge a_2.s) \vee a_2.f} {({\bar a}_f \vee a_1.s) \wedge a_2.s}}{\ef{e_1 \rhd (({\bar a}_f \vee a_1.s) \wedge e_2)}}
  }
}

It is useful to recognize how $\textcolor{blue}{a_1}$ and $\textcolor{blue}{a_f}$ act as filters on $\textcolor{blue}{a_2.s}$ and $\textcolor{red}{e_2}$, which, \eg, allows
us to recover freshness. If $t_1$ is fresh and $\textcolor{blue}{a_1.f = \btrueA}$, $t_2$ will have $\textcolor{blue}{a_2.s = \btrueA}$. But if $\textcolor{blue}{a_1.s}$ and $\textcolor{blue}{a_f = \bfalseA}$,
the result for the entire let will be fresh only, not stored. Thus, \eg, $\tlet x {\tref \tfalse} x$ has type $\TRef\ {\avc \btrueA \bfalseA}\ \bfalseE $.

\bfparagraph{Strengthening: Safe Purity Downcast Based on Ambient Ability}

This style of reasoning goes even further. We use the notation $\hasTypeAEC {\GA \ab{a_f}} {t} {T} {a} {e}$ from \secref{sec:ability} and
establish \lemref{lem:hast_strengthen},
which allows us to strengthen the
ability and effect typing based on the term's ambient ability $\textcolor{blue}{a_f}$:
\begin{lemma}[Strengthening]\label{lem:hast_strengthen} The following typing rule is admissible:
\infrule[T-Strengthen]{
  \hasTypeAEC {\GA \ab{a_f}} {t} {T} {a} {e}
}{
  \hasTypeAEC {\Gamma} {t} {T} {\av {a.f} {{\bar a}_f \wedge a.s}} {{\bar a}_f \wedge e}
}
\end{lemma}
The proof crucially relies on the filtering behavior of the \textsc{T-Abs-AE} and \textsc{T-App-AE} rules.

\bfparagraph{Syntactic Purity}

Which expressions are syntactically pure? In $\lambda_{ae}$, 
a term assigned any ability qualifier $\textcolor{blue}{\av f s}$ and effect $\btrueE$ (after strengthening) is impure, since effect $\btrueE$ means 
that evaluating the term may induce observable effects.
A term assigned qualifier \textcolor{blue}{$\av \btrue s$} $\textcolor{red}{e}$ is also impure, as it yields a freshly allocated location, which may be observed by the execution context.
A term assigned qualifier \textcolor{blue}{$\av \bfalse \btrue$} $\bfalseE$ may return
pre-existing resources but no freshly allocated locations, and it has no observable effects ($\bfalseE$). This is still pure! Thus, pure expressions are characterized (after strengthening) by the typing $\hasTypeAEC {\Gamma} {t} {T} {\av \bot s} \bfalseE$.

\subsection{Encoding of Pure Effect and Pure Ability System}\label{sec:encoding}

\begin{figure}[t]\small
    \begin{mdframed}[leftmargin=2pt, rightmargin=2pt]
        \typicallabel{TR-A-AE}
        \judgement{Type Encoding of $\lambda_e$}{$\BOX{\trTE{\TE}}$}\\\vspace{-2ex}
        \[
            \begin{array}{l@{\quad}l@{\quad}l@{\quad}l@{\quad}l@{\quad}l@{\quad}l@{\quad}l@{\quad}l@{\quad}l@{\quad}l@{\quad}l@{\quad}l}
            \trTE{\TBoolE} & = & \TBoolAE & & &
            \trTE{\TRefE} & = & \TRefAE \\ 
            \trTE{\TFunE{\TE_1}{\TE_2}{e}} & = & \multicolumn{6}{l}{\!\!\!\!\TFunAE{\trTE{\TE_1}}{{\av{\btrueA}{\btrueA}}}{\trTE{\TE_2}}{{\av{\btrueA}{e}}}{e}}
        \end{array}%
        \]
        \\[-2ex]
        \judgement{Type Encoding of $\lambda_a$}{$\BOX{\trTA{\TA}}$}\\\vspace{-2ex}
        \[
            \begin{array}{l@{\quad}l@{\quad}l@{\quad}l@{\quad}l@{\quad}l@{\quad}l@{\quad}l@{\quad}l@{\quad}l@{\quad}l@{\quad}l@{\quad}l}
            \trTA{\TBoolA} & = & \TBoolAE & & &
            \trTA{\TRefA} & = & \TRefAE \\ 
            \trTA{\TFunA{\TA_1}{a_1}{\TA_2}{a_2}} & = & \multicolumn{6}{l}{\!\!\!\!\TFunAE{\trTA{\TA_1}}{{\av{a_1}{a_1}}}{\trTA{\TA_2}}{{\av{a_2}{a_2}}}{\btrueE}}
        \end{array}%
        \]
    \end{mdframed}   
    \caption{Type translation from $\lambda_e$ and  $\lambda_a$ to $\lambda_{ae}$ respectively.}
    \label{fig:tr} 
\end{figure} Now we show that $\lambdaae$ can encode the effect system $\lambdae$ and the ability system $\lambdaa$ in \figref{fig:tr}.
The base types are encoded directly. 
For function types encoding of $\lambdae$, as the argument and return types in $\lambdae$ do not have information about ability, 
the encoding conservatively assumes the $\btrueA$ for the fresh and stored components
in their translated types.
For function types encoding of $\lambdaa$, %
as $\lambdaa$ does not have information on the function's latent effects, the encoding conservatively assigns the effect $\btrueE$,
relying on the fact that it will be only considered relative to the ability of the function.

We define $\trTE{\GE}$ and $\trTA{\GA}$ that translate the respective source typing environments, 
to the target ones by mapping each type according to the transition rules defined in \figref{fig:tr}.
The following theorems show that the encodings are sound and preserve syntactic purity. %

\begin{theorem}\label{thm:e_to_ae}
  If $\hasTypeE{\GE}{t}{\TE}{e}$
  then $\hasTypeAEC{\trTE{\GE}}{t}{\trTE{T}}{\av{e}{\btrueA}}{e}$.
\end{theorem}
\begin{theorem}\label{thm:a_to_ae}
If $\hasTypeA{\GA \ab{a_f}}{t}{\TA}{a}$
then $\hasTypeAEC{\trTA{\GA}}{t}{\trTA{T}}{\av{a}{{\bar a}_f \wedge a}}{\ef{{\bar a}_f}}$.
\end{theorem}

\begin{corollary}\label{thm:ae_expressiveness}
The syntactically pure terms of $\lambdae$ and $\lambdaa$ are syntactically pure in $\lambdaae$.
\end{corollary}

The following lemma shows that the translation of context typing for $\lambdae$ is sound: any well-typed context in $\lambdae$ remains well-typed in $\lambdaae$ after the translation.
\begin{lemma}\label{lem:tr_ce}
    if $C: (\Gamma; T\ \ef{e}) \carrow (\Gamma'; T'\ \ef{e'}\ \flt') $
    then 
    \[ 
    C: (\trTE{\Gamma\ \btrueA}; \trTE{T}\ \av{\ab{e}}{\btrueA}\ \ef{e}) \carrow  (\trTE{\Gamma'\ \flt'}; \trTE{T'}\ \av{\ab{e'}}{\btrueA}\ \ef{e'}).  
    \]  
\end{lemma}   

The following lemma shows that the translation of context typing for $\lambdaa$ is sound: any well-typed context in $\lambdaa$ remains well-typed in $\lambdaae$ after the translation.
\begin{lemma}\label{lem:tr_ca}
    if $C: (\Gamma\ \ab{a_{\flt}}; T\ \ab{a}) \carrow (\Gamma'\ \flt'; T'\ \ab{a'}) $
    then 
    \[ 
    C: (\trTA{\Gamma\ \ab{a_{\flt}}}; \trTA{T}\ \avc{a}{a}\ \btrueE) \carrow  (\trTA{\Gamma'\ \flt'}; \trTA{T'}\ \avc{a'}{a'}\ \btrueE).  
    \]  
\end{lemma}

\subsection{Discussion}\label{sec:discuss_tae}
\begin{figure}[t]\small
  \begin{mdframed}[leftmargin=2pt, rightmargin=2pt]
    \typicallabel{T-False-E}
    \judgement{Syntax}{$\BOX{t ::= ...}$}\vspace{-2ex}
    \begin{minipage}[t]{\linewidth}
      \[\begin{array}{l@{\quad}l@{\quad}l@{\quad}l@{\quad}l@{\quad}l@{\quad}l@{\quad}l@{\quad}l@{\quad}l@{\quad}l@{\quad}l@{\quad}l}
        t & := & \cdots \mid \tnil \mid \tcons{t_1}{t_2} \mid \tfold{(x, y => t_1)}{t_2}{t_3} & \text{Terms}         \\
        v & := & \cdots \mid  [\overline{v}] & \text{Values}              \\
        T & := & \cdots \mid \TListAE{T} & \text{Types}              \\
      \end{array}\]\\ 
    \end{minipage}%
    \vspace{-4ex}\\

    \judgement{Term Typing}{$\BOX{\hasTypeAEC \Gamma t T {a} e}$}\vspace{-4ex}
   \begin{multicols}{2}
     
    \infrule[T-Nil-AE]{
    }{
      \hasTypeAEC \Gamma {\tnil} {\TListAE{T}} {\avbot} {\bfalseE}
    }
    
    \infrule[T-Cons-AE]{
      \hasTypeAEC \Gamma {t_1} {T} {a_1} {e_1} \\
      \hasTypeAEC \Gamma {t_2} {\TListAE{T}} {a_2} {e_2}
    }{
      \hasTypeAEC \Gamma {\tcons{t_1}{t_2}} {\TListAE{T}} {a_1 \vee a_2} {e_1 \rhd e_2}
    }
    
    \infrule[T-Fold-AE]{
      \hasTypeAEC {\Gamma, x : T_2\ \ab{a_2}, y: T_1\ \ab{a_1}} {t_1} {T_2} {a_2} {e_2}  \\
      \hasTypeAEC {\Gamma} {t_2} {T_2} {a_2} {e_z}  \\ %
      \hasTypeAEC {\Gamma} {t_3} {\TListAE{T_1}} {a_1} {e_1} %
    }{
      \hasTypeAEC {\Gamma} {\tfold{(x, y => t_1)}{t_2}{t_3}} {T_2} {a_2} {e_z\rhd e_1\rhd e_2}
    }
   \end{multicols}
  \end{mdframed}
  \caption{Syntax and typing rules for lists (operational semantics elided). Note that the fresh component of ability in rule \textsc{T-Fold-AE} is always $\bfalseA$. 
  We write \textcolor{blue}{$a_1 \vee a_2$} to mean component-wise disjunction of $a_1$ and $a_2$.} %
  \label{fig:list}
\end{figure} %
\begin{figure}[t]\small
\begin{mdframed}
 \centering
 \begin{lstlisting}
 let x = true in                  // : Bool $\avbot$      in context [ x: Bool $\avbot$ ]
 let y = ref(true) in             // : Ref $\av{\bfalseA}{\btrueA}$  in context [ y: Ref $\av{\btrueA}{\bfalseA}$, x: Bool $\avbot$ ]   
 let a = ref(false) in            // : Ref $\av{\bfalseA}{\btrueA}$  in context [ a: Ref $\av{\btrueA}{\bfalseA}$, y: Ref $\av{\btrueA}{\bfalseA}$, 
                                  // :                         x: Bool $\avbot$ ]
 def id (x: Bool) = { x }         // : (Bool $\avbot$      =>$^{\bfalseE}$ Bool $\avbot$) $\avbot$  
 def idr (x: Ref) = { x }         // : (Ref $\av{\btrueA}{\bfalseA}$   =>$^{\bfalseE}$ Ref $\av{\bfalseA}{\btrueA}$) $\avbot$ 
 def alloc (x: Bool) = { ref x }  // : (Bool $\avbot$      =>$^{\bfalseE}$ Ref $\av{\btrueA}{\bfalseA}$) $\avbot$ 
 def leak: (x: Bool) = { a }      // : (Bool $\avbot$      =>$^{\bfalseE}$ Ref $\av{\bfalseA}{\btrueA}$) $\av{\bfalseA}{\btrueA}$ 
 def use (x: Ref) = { !a }        // : (Ref $\av{\btrueA}{\bfalseA}$    =>$^{\btrueE}$ Bool $\avbot$) $\av{\bfalseA}{\btrueA}$ 
 def usearg (x: Ref) = { !x }     // : (Ref $\av{\bfalseA}{\btrueA}$       =>$^{\btrueE}$ Bool $\avbot$) $\avbot$ 
 def mention (x: Ref) = { true }  // : (Ref $\av{\bfalseA}{\btrueA}$  =>$^{\bfalseE}$ Bool $\avbot$) $\avbot$ 
 def capture (x: Ref) = { $\lambda$ y:Ref. !a } 
                                  // : (Ref $\av{\btrueA}{\bfalseA}$  =>$^{\bfalseE}$ ((Ref $\av{\btrueA}{\bfalseA}$  =>$^{\btrueE}$ Bool $\avbot$) $\av{\bfalseA}{\btrueA}$)) $\av{\bfalseA}{\btrueA}$  
 id(x)                            // : Bool $\avbot$ $\bfalseE$
 idr(a)                           // : Ref $\av{\bfalseA}{\btrueA}$ $\bfalseE$
 use(y)                           // : Bool $\avbot$ $\btrueE$
 usearg(a)                        // : Bool $\avbot$ $\btrueE$
 mention(a)                       // : Ref $\av{\bfalseA}{\btrueA}$ $\bfalseE$
 \end{lstlisting}    
\end{mdframed} 
\caption{Selected examples in Figure 2 of the main paper that are annotated with abilities and effects.}
\label{fig:purity_with_types}
\end{figure}     
\bfparagraph{Case Study: Lists with Fold and Map} 
Following Sections~\ref{sec:discuss_effs} and~\ref{sec:diss_ability}, we 
consider once more higher-order functions such as |map|.
In \figref{fig:list}, we extend $\lambdaae$ with list constructors: 
$\tnil$ denotes the empty list, 
and $\tcons{t_1}{t_2}$ constructs a list by prepending $t_1$ to the list $t_2$.
The extension also includes the fold operation: $\tfold{(x, y => t_1)}{t_2}{t_3}$,
which fold the list $t_3$ using the function $t_1$ and the initial accumulator $t_2$.
The semantics of $\tnil$ and $\tcons{t_1}{t_2}$ are straightforward.
For $\tfold{(x, y => t_1)}{t_2}{t_3}$, the semantics first evaluate $t_2$ to obtain the initial accumulator,
then evaluate $t_3$ to obtain a list of values,
and finally, for each of the element in the list, it evaluates $t_1$ 
in the context extended with the current accumulator $x$ and an element $y$ in the list. 
The end-to-end mechanization is developed in Rocq.

In this extension, the function |map| can be encoded and typed as: 
\[\small
\begin{array}{rcl}
map & \DEF & \lambda f.\ \lambda ls.\ \tfold{(x, y => {\tcons {(\tapp f x)} y})}{\tnil}{ls}\\
map & : & \ \  (\TFunAE {T_1} {\avbot} {T_2} {\avbot} {\btrueE})\ \avtop\ \quad \stackrel{\textcolor{red}{\bfalseE}}{\to} \quad 
      (\TFunAE {\TListAE{T_1}} {\avbot} {\TListAE{T_2}} {\avbot} {\btrueE})\ {\avc \bfalseA \btrueA} \\
\end{array}
\]
With the result qualifier ${\avc \bfalseA \btrueA}$, a partial application 
|map(f)| never returns a fresh resource (and is thus pure) unless |f| is a fresh resource. 
A weaker version with $\avtop$ would make partial applications |map(f)| appear impure.
Let us consider a concrete example: 
assume a list of Booleans $\texttt{ls} := [\ttrue, \tfalse, \ttrue]$ 
with type $\TListAE{\TBoolAE}\ \avbot\ \bfalseE$, and a pure function $\texttt{neg} := \lambda x. \neg x$ that negates its argument value,
with the pure type $(\TFunAE {\TBoolAE}{\avbot}{\TBoolAE}{\avbot}{\bfalseE})\ \avbot\ \bfalseE$.
Based on rule \textsc{T-App-AE}, the partial application |map(neg)| has type $(\TFunAE {\TListAE \TBoolAE}{\avbot}{\TListAE \TBoolAE}{\avbot}{\bfalseE})\ {\avc \bfalseA \btrueA}\ \bfalseE$ which is pure, and could be further strengthened to $(\TFunAE {\TListAE \TBoolAE}{\avbot}{\TListAE \TBoolAE}{\avbot}{\bfalseE})\ \avbot\ \bfalseE$. 
Applying the result to |ls| yields |map(neg)(ls)| with type
${\TListAE \TBoolAE}\ \avbot\ \btrueE$ according to \textsc{T-App-AE}, but can be
strengthened to the pure type ${\TListAE \TBoolAE}\ \avbot\ \bfalseE$.
Thus, we obtain the desired lightweight effect polymorphism as in $\lambdaa$, but
coupled with a use-mention distinction as in $\lambdae$ (partial applications
remain pure as well).

\bfparagraph{Other Examples}
\figref{fig:purity_with_types} shows selected examples from \figref{fig:intuitionExamples} annotated with abilities and effects in $\lambdaae$.
The function |alloc| creates a new reference initialized with the given Boolean argument. 
The allocation does not induce observable effects, but the resulting reference carries an ability $\av{\btrueA}{\bfalseA}$,
meaning that the return value is fresh.
The function |leak| returns the reference captured from the environment, which is not fresh. Thus, its ability is $\av{\bfalseA}{\btrueA}$.
The functions |use| and |usearg| induce effects on the captured reference from the environment and its argument respectively.
Thus, their effects are $\btrueE$.
Note that the ability of |use|'s argument can be $\avbot$, as the function itself does not use its argument.
But, |use(a)| becomes untypeable.
In contrast, the function |mention| does not use its argument's ability or capture anything from the environment; 
hence, its effect is $\bfalseE$. 
In addition, the functions |leak| and |use| capture the ability carried by the reference |a| from the environment, thus their abilities are $\av{\bfalseA}{\btrueA}$; 
others are just $\avbot$.

\bfparagraph{Conclusion}
In terms of expressiveness, $\lambdaae$ subsumes both $\lambdae$
and $\lambdaa$ in their ability to identify pure programs. 
Corollary~\ref{thm:ae_expressiveness} establishes that all syntactically pure
programs of $\lambdae$ and $\lambdaa$ are also pure in $\lambdaae$. This
includes all the challenge examples discussed in 
Sections~\ref{sec:discuss_effs} and~\ref{sec:diss_ability}.

\section{Observational Equivalence and Purity via Binary Logical Relations for $\lambdaae$}\label{sec:lr}

In this section, we present the details of the earlier theorems,
which are proved using binary logical relations that show semantically equivalent values and terms are logically related. 

\subsection{Observational Equivalence}
\begin{figure}[t]\small
    \begin{mdframed}
        \judgement{Context for Observational Equivalence}{}
         \[\begin{array}{l@{\ \ }c@{\ \ }l@{\qquad\qquad\ }l@{\ \ }c@{\ \ }l}
				{C} & ::= & \square \mid \tapp{C}{t} \mid  \tapp{t}{C} \mid  \tabs{C} \mid \tref{C} \mid\ \tget{C} \mid \tput{C}{t} \mid \tput{t}{C} \mid \tseqb{t_1}{C} \mid \tseqb{C}{t_2} &  &  & \\
			\end{array}\]
            \judgement{Context Typing Rules}{
                \BOX{C : (\Gamma; T\ \ab{a}\ \ef{e}\ \ab{a_\flt}) \carrow (\Gamma'; T'\ \ab{a'}\ \ef{e'}\ \flt')}
            }
            
            \infax[c-$\square$]{
                \square: (\Gamma; T\ \ab{a}\ \ef{e}\ \ab{a_\flt}) \carrow (\Gamma; T\ \ab{a}\ \ef{e}\ \flt)
            }

            \infrule[c-$\lambda$]{
               C: (\Gamma; T\ \ab{a}\ \ef{e}\ \ab{a_\flt}) \carrow ((\G', x: T_1\ \ab{a_1}); T_2\ \ab{a_2}\ \ef{e_2}\ \flt') \andalso
               \envCapA{\Gamma'}{\flt' \setminus x}{a_f} \\
               \textcolor{blue}{a_f'} = \avc{\bfalse}{(a_f.s \vee a_{\flt}.s) \wedge (a_2.s \vee e_2)}
            }{
              \tabs{C}:(\Gamma; T\ \ab{a}\ \ef{e}\ \ab{a_\flt}) \carrow {\Gamma'; (\TFunAE{T_1}{a_1}{T_2}{a_2}{e_2})\ \ab{a_f'}\ \bfalseE \ (\flt' \setminus x)}
            }

            \infrule[c-app-1]{
              C: (\Gamma; T\ \ab{a}\ \ef{e}\ \ab{a_\flt}) \carrow (\Gamma'; (\TFunAE{T_1}{a_1}{T_2}{a_2}{e})\ \ab{a_f}\ \ef{e_f}\ \flt') \andalso
              \hasTypeAE{\Gamma'}{t}{T_1}{\ab{a_1}}{\ef{e_1}} 
            }{
              \tapp{C}{t}: (\Gamma; T\ \ab{a}\ \ef{e}\ \ab{a_\flt}) \carrow (\Gamma'; T_2\ \\
               {\avc 
              {a_2.f \vee a_2.s \wedge (a_f.f \vee a_1.f)}
              {a_2.s \wedge (a_f.s \vee a_1.s)}}\
              \ef{e_f \rhd e_1 \rhd (e \wedge (a_f.s \vee a_1.s))}\ \flt')
            }

            \infrule[c-app-2]{
             \hasTypeAE{\Gamma'}{t}{\TFunAE{T_1}{a_1}{T_2}{a_2}{e}}{\ab{a_f}}{\ef{e_f}} \andalso 
             C: (\Gamma; T\ \ab{a}\ \ef{e}\ \ab{a_\flt}) \carrow (\Gamma'; T_1\ \ab{a_1}\ \ef{e_1}\ \flt_1) 
            }{
              \tapp{t}{C}: (\Gamma; T\ \ab{a}\ \ef{e}\ \ab{a_\flt}) \carrow (\Gamma'; T_2\ \\
               {\avc 
              {a_2.f \vee a_2.s \wedge (a_f.f \vee a_1.f)}
              {a_2.s \wedge (a_f.s \vee a_1.s)}}\
              \ef{e_f \rhd e_1 \rhd (e \wedge (a_f.s \vee a_1.s))}\ \fv{t})
            }

            \infrule[c-ref]{                 
               C: (\Gamma; T\ \ab{a}\ \ef{e}\ \ab{a_\flt}) \carrow (\Gamma'; \TBoolA\ \ab{a'}\ \ef{e'}\ \flt')
            }{
              \tref{C}:(\Gamma; T\ \ab{a}\ \ef{e}\ \ab{a_\flt}) \carrow (\Gamma'; \TRefA\ {\avc \btrue \bfalse}\ \ef{e'}\ \flt')   
            }

            \infrule[c-get]{
                C:(\Gamma; T\ \ab{a}\ \ef{e}\ \ab{a_\flt}) \carrow (\Gamma'; \TRefA\ \ab{a'}\ \ef{e'}\ \flt')
            }{
                \tget{C}: (\Gamma; T\ \ab{a}\ \ef{e}\ \ab{a_\flt}) \carrow (\Gamma'; \TBoolA; \avbot \ \ef{e' \rhd a'.s}\ \flt')
            }

            \infrule[c-put-1]{
               C: (\Gamma; T\ \ab{a}\ \ef{e}\ \ab{a_\flt}) \carrow (\Gamma'; \TRefA\ \ab{a_1}\ \ef{e_1} \ \flt_1) \andalso 
               \hasTypeAE {\Gamma}{t}{\TBoolA}{\ab{a_2}}{\ef{e_2}}
            }{
                \tput{C}{t}: (\Gamma; T\ \ab{a}\ \ef{e}\ \ab{a_\flt}) \carrow (\Gamma'; \TBoolA\ \avbot \ \ef{e_1 \rhd e_2 \rhd a_1.s}\ \flt_1)
            }

            \infrule[c-put-2]{
               \hasTypeAE {\Gamma}{t}{\TRefA}{\ab{a_1}}{\ef{e_1}} \andalso
               C: (\Gamma; T\ \ab{a}\ \ef{e}\ \ab{a_\flt}) \carrow (\Gamma'; \TBoolA\ \ab{a_2}\ \ef{e_2}\ \flt_2) 
            }{
                \tput{t}{C}: (\Gamma; T\ \ab{a}\ \ef{e}\ \ab{a_\flt}) \carrow (\Gamma'; \TBoolA\ \avbot \ \ef{e_1 \rhd e_2 \rhd a_1.s} \ \flt_2)
            }

           \infrule[c-bin-1]{
              \hasTypeAE{\Gamma}{t}{\TBoolA}{\ab{a_1}}{\ef{e_1}} \andalso
              C; (\Gamma; T\ \ab{a}\ \ef{e}\ \ab{a_\flt}) \carrow (\Gamma'; \TBoolA\ \ab{a_2}\ \ef{e_2}\ \flt_2)
            }{
             \tseqb{t}{C}: (\Gamma; T\ \ab{a}\ \ef{e}\ \ab{a_\flt}) \carrow (\Gamma'; \TBoolA\ \avbot \ \ef{e_1 \rhd e_2} \ \flt_2)
            }

            \infrule[c-bin-2]{
                C; (\Gamma; T\ \ab{a}\ \ef{e}\ \ab{a_\flt}) \carrow (\Gamma'; \TBoolA\ \ab{a_1}\ \ef{e_1}\ \flt_1) \andalso
                \hasTypeAE{\Gamma}{t}{\TBoolA}{\ab{a_2}}{\ef{e_2}}
              }{
               \tseqb{C}{t}: (\Gamma; T\ \ab{a}\ \ef{e}\ \ab{a_\flt}) \carrow (\Gamma'; \TBoolA\ \avbot \ \ef{e_1\rhd e_2}\ \flt_1)
              }
            
            \infrule[c-sub]{
              C:(\Gamma; T\ \ab{a}\ \ef{e}\ \ab{a_\flt}) \carrow (\Gamma'; T'\ \ab{a'}\ \ef{e'} \ \flt') \andalso
              T' <: T'' \andalso
              \ab{a' <: a''} \andalso
              \ef{e' <: e''}
            }{
              C:(\Gamma; T\ \ab{a}\ \ef{e}\ \ab{a_\flt}) \carrow (\Gamma'; T''\ \ab{a''}\ \ef{e''} \ \flt')
            }

    \end{mdframed}
    \caption{Context typing rules for $\lambdaae$. The definition of context is repeated for easy reference.}
    \label{fig:context}
\end{figure}
 We write $\ctxEquivAE{\G} {t_1} {t_2} {T} {a} {e}$ to mean that 
programs $t_1$ and $t_2$ are \emph{observationally equivalent} at type $T\ \ab{a}\ \ef{e}$, 
meaning that they exhibit indistinguishable behavior in all program contexts $C$ with a hole of type $T\ \ab{a}\ \ef{e}$.
That is to say if $C[t_1]$ exhibits some observable behavior, then so does $C[t_2]$.

Following prior work~\cite{logical-approach,DBLP:conf/popl/AhmedDR09,bao2025logrel}, we define
a judgement for logical equivalence using binary logical relations,
written as $\semTypeAE{\G} {t_1} {t_2} {T} {a} {e}$. 
It means that terms $t_1$ and $t_2$ are logically related at type $T$, ability $\AB{a}$ and effect $\EF{e}$ under $\Gamma$.

Differing from reduction contexts, 
contexts $C$ for reasoning about equivalence allow a ``hole'' to appear in any place.
We write $C: (\G\ \ab{a_{\flt}}; T\ \textcolor{blue}{a}\ \textcolor{red}{e}) \carrow (\G'\ \flt'; T'\ \textcolor{blue}{a'} \textcolor{red}{e'})$ to 
mean that the context $C$ is a program, %
which has the type $T'\ \textcolor{blue}{a'}\ \textcolor{red}{e'} $ (closed under $\G'$) and whose free variables are in $\flt'$, 
with a hole that can be filled 
with any program $t$ such that $\G\, \ab{a_{\flt}} \vdash T \ \ab{a}\ \ef{e}$.
The typing rules for well-typed contexts are shown in \figref{fig:context}, and imply that if
$\hasTypeAEC{\G}{t}{T}{a}{e}$,
$\flt' = \fv{C[t]}$
and $C:  (\G\ \ab{a_{\flt}}; T\ \textcolor{blue}{a}\ \textcolor{red}{e})\carrow (\G'\ \flt'; T'\ \textcolor{blue}{a'}\ \textcolor{red}{e'})$ hold, 
then $\hasTypeAEC{\G'} {C[t]} {T'} {a'} {e'}$.

We adapt the standard definition of observational equivalence~\cite{DBLP:conf/popl/AhmedDR09}
to use a Boolean context, since programs typed with our system always terminate. 
\begin{definition}[Observational Equivalence]\label{def:standard_equiv} Term $t_1$ is \emph{observationally equivalent} to $t_2$,
  written $\ctxEquivAE{\G}{t_1}{t_2}{T}{a}{e}$, 
  if $\hasTypeAEC{\G\ \ab{a_{\flt}}} {t_1} {T} {a} {e}$, and 
  $\hasTypeAEC{\G\ \ab{a_{\flt}}} {t_2} {T} {a} {e}$, and 
 $\forall \, C: (\G\ \ab{a_{\flt}}; T\ \textcolor{blue}{a}\ \textcolor{red}{e}) \carrow (\emptyset\ \qempty; \TBoolA\ \avc{\bfalseA}{\btrueA} \btrueE). \
        C[t_1] \ \simeq \ C[t_2]$.
\end{definition}

\subsection{The World Model}\label{sec:world}

We apply Kripke logical relations to our system.
Our logical relations are indexed by types and store layouts via \emph{worlds},
generalized store typings relating two stores.
The invariant that all allocated locations hold well-typed values \wrt the world must hold in the pre-state and be re-established in the post-state of a computation.
This invariant must hold in future worlds where more locations may be allocated, which is commonly referred to as \emph{monotonicity}.
The notion of world is defined as: 
\begin{definition}[World]\label{def:world} A world $\W$ is a triple of $(\Sigma_1, \Sigma_2, f)$, where
    \begin{itemize}
        \item $\Sigma_1$ and $\Sigma_2$ are store typings defined in \figref{fig:world},

        \item $ f \subseteq (\DOM(\Sigma_1) \times \DOM(\Sigma_2))$ is a partial bijection.
    \end{itemize}
\end{definition}
A world defines relational stores.
A partial bijection $f$ means that if $f(\ell_1, \ell_2)$ and $f(\ell_3, \ell_4)$, 
then $\ell_1 = \ell_3$ if and only if $\ell_2 = \ell_4$.
It captures the fact that a relation holds under permutation of store locations.
If $\W = (\Sigma_1, \Sigma_2, f)$ is a world, we refer to its components as follows: 
$ \W(\ell_1, \ell_2) = (\ell_1, \ell_2) \in f$ if $\ell_1 \in \DOM(\Sigma_1) \text{ and } \ell_2 \in \DOM(\Sigma_2)$ and when $f$ defined; 
otherwise $\bot$.
We write $\DOM_1(\W)$ and $\DOM_2(\W)$ to refer the first and second components in the world $\W$, 
\ie, $\DOM_1(\W) = \DOM(\Sigma_1)$ and $\DOM_2(\W) = \DOM(\Sigma_2)$. 
Let $\W$ and $\W'$ be two worlds. We write $\DOM(\W) \subseteq \DOM(\W')$ to mean 
$\DOM_1(\W) \subseteq \DOM_1(\W') \, \land \, \DOM_2(\W) \subseteq \DOM_2(\W')$.
We write $\W_f$ to mean the third component of the world $\W$.

If $\W$ and $\W'$ are worlds, such that
$
    \DOM_1(\W) \cap \DOM_1(\W') = \DOM_2(\W) \cap \DOM_2(\W') = \emptyset
$,
then $\W$ and $\W'$ are called disjoint, and we write $\W \extends \W'$ to mean extending $\W$ with a disjoint world $\W'$. 
Let $\sigma_1$ and $\sigma_2$ be two stores. 
We write $\st{\sigma_1}{\sigma_2}{\W}{L_1}{L_2}$ to mean the stores are well-defined \wrt the world $\W$ at a pair of locations $(L_1, L_2)$, 
which is formally defined in \figref{fig:world}.

\begin{figure}[t]\small
  \begin{mdframed}[leftmargin=2pt, rightmargin=2pt]
  \judgement{Store Typing, Relational Worlds, Well-Defined Stores}{$\BOX{\lambdaae}$}\vspace{-2ex}
  \begin{mathpar}
    \begin{array}{@{\hspace{-2ex}}r@{\hspace{1ex}}l@{\hspace{1ex}} l}
        \Sigma &:= &\qbot \mid \Sigma, \ell: \{ \ttrue, \tfalse \} \\

        \stcp{\W}{\W'}{\lsv{1}}{\lsv{2}} & = & (\forall \, \ell_1 \in \lsv{1}, \ell_2 \in \lsv{2}. \, \W_f(\ell_1, \ell_2) \Rightarrow \W'_f(\ell_1, \ell_2)) \\ 
        \st{\sigma_1}{\sigma_2}{\W}{\lsv{1}}{\lsv{2}}  & \DEF & \sigma_1: \W_1 \, \land \, \sigma_2: \W_2 \, \land \, (\forall (\ell_1, \ell_2) \in \W. \, \ell_1 \in \lsv{1} \, \land \, \ell_2 \in \lsv{2} \, \Rightarrow  \, (\exists \, b.\, \sigma_1(\ell_1) = \sigma_2(\ell_2) = b )) \\
    \end{array}
\end{mathpar}%
\judgement{\textsf{\textbf{Reachable Locations}}\phantom{lifiers}}{}\vspace{0ex}\\
$
  \begin{array}{l@{\ \,}c@{\ \,}l@{\qquad\qquad\qquad\qquad\ \ }r}
    \vallocs{b} = \qbot    \qquad
    \vallocs{\ell} = \{\ell\}   \qquad
    \vallocs{\cll{H}{\lambda x.t}} = \explocs{\lambda x.t}{H} \qquad
    \varslocs{q}{H} = \bigcup_{x \in q} L(H(x))
  \end{array}
$    
\end{mdframed}
\caption{Definitions of store typing, relational words and well-defined stores \wrt a world; and computing reachable locations from a given value.}
\label{fig:world}
\end{figure} 

\bfparagraph{Reachable Locations}
Following \citet{bao2025logrel}'s work,
our logical relations need to know the reachable locations from a given value $v$, 
computed by the function $\vallocs{v}$ (\figref{fig:world}).
Boolean values reach the empty set of locations, and a location $\ell$ reaches itself.
A function value $\tabs{t}$ reaches the locations from the values retrieved from its free variables in the value environment.

\bfparagraph{Relational Worlds} 
Instead of using the standard definition of world extension,
we adapt the relational worlds from \citet{bao2025logrel}'s work, 
which leverages reachable locations to achieve a degree of localization.
In \figref{fig:world}, relational worlds,
written as $\stcp{\W}{\W'}{\lsv{1}}{\lsv{2}}$,
parameterize the definition of world extension with related reachable locations $\lsv{1}$ and $\lsv{2}$.
The standard definition of store typing extension $\W \sqsubseteq \W'$ can be defined as $\stcp{\W}{\W'}{\DOM_1(\W)}{\DOM_2(\W)}$.
The relational worlds are used in reasoning about functions. See detailed explanation in \secref{sec:valt}.

\subsection{Use vs Mention Aware Binary Logical Relations}\label{sec:lr_def}
We first introduce the interpretation of typing contexts, as premise of semantic typing.
\begin{definition}[Semantic Typing Context]\label{def:sem_ctx} The definition of typing context interpretation is:
\[
\small
\begin{array}{@{\hspace{2ex}}r@{\hspace{1ex}}l@{\hspace{1ex}} l}
       \VG{\G \ \flt\ u}  & = &  \{(\W, \VE, \R) \mid \DOM(\VE_i) =  \DOM(\R_i) = \DOM(\G) \, \wedge \, (\forall x, T, \ab{a}. \, \G (x) = T \ \ab{a} \Rightarrow\ \\
                       &   &  ~~~~ \exists\ u_x, L_1, L_2.\, u_x = (\ab{\neg \bar{a}} \vee u)\, \wedge\, (x \in \flt \Rightarrow (\W, \VE_1(x), \VE_2(x), u_x, L_1, L_2) \in \VT{T}) \, \land \, \\
                       &   & \qquad\qquad\!\! ((\ab{\neg \bar{a}} \vee u) \Rightarrow \R_1(x) = L_1 \wedge \R_2(x) = L_2)\ \wedge \, ((\ab{\neg \bar{a}} \vee \neg u) \Rightarrow (L_1 = \emptyset \, \wedge \, L_2 = \emptyset)))\}. 
\end{array}
\]
\end{definition}
\noindent A semantic typing context is defined as a set of tuples of the form $(\W, \VE, \R)$, where $\W$ is a world (\defref{def:world}).
The component $\VE$ ranges over a pair of relational value environments that are finite maps from variables $x$ to pairs of values $(v_1, v_2)$.
If $\VE(x) = (v_1, v_2)$, then $\VE_1(x)$ denotes $v_1$ and $\VE_2(x)$ denotes $v_2$.
We write $\DOM(\VE_1)$ and $\DOM(\VE_2)$ to mean the domain of the first resp.\ second value environment. 
The component $\R$ ranges over a pair of relational reachability environments 
that are finite maps from variables $x$ to pairs of locations $(L_1, L_2)$, which are reachable from their respective values, \ie, $\VE_1(x)$ and $\VE_2(x)$.
The domains of $\R_1$ and $\R_2$ are defined similarly to those of $\VE$.

The reachability information defined in $\R$ 
are refined \wrt a reasoning mode, which allows reasoning to 
distinguish whether external resource is actually used or only mentioned in two pairs of terms, 
\wrt the effect qualifier. 
The mode is determined by the semantic typing judgement.

\begin{definition}[Semantic Typing Judgment]\label{def:sem_type} The semantic typing judgement is defined as:
\[
\small
\begin{array}{@{\hspace{2ex}}r@{\hspace{1ex}}l@{\hspace{1ex}} l}
\semTypeAE{\G}{t_1}{t_2}{T}{a}{e} \DEF & \forall\, u. (\ef{e} \Rightarrow u) \Rightarrow \, (\forall\, \W, \VE, \R.  \\
 & (\W, \VE, \R)\in \VG{\G \ (\fv{t_1} \cup \fv{t_2})\ u}. \ (\W, \VE, t_1, t_2, u) \in \ETAE{T}{a}{e}).
\end{array}
\]
\end{definition}
\noindent 
Here, the parameter $u$ means that the current reasoning is in a use mode if it is \texttt{true}; 
otherwise it is in a mention mode, meaning that the current reasoning does not require the knowledge about any external resource.
Here, we write $\ef{e} \Rightarrow u$ to mean $\ef{e} = \btrueE \Rightarrow u = \texttt{true}$.
The condition dictates that the reasoning must be in a use mode %
whenever the computation induces observable effects. %
\bfparagraph{Example} \figref{fig:purity_with_types} shows the examples annotated with abilities and effects. 
The program |mention(a)| can be reasoned in either mode, (\ie, $u = \texttt{true}$ or $u = \texttt{false}$),
as they don't have observable effects. %
In contrast, the programs |use(y)| and |usearg(a)| can only be reasoned in the use mode (\ie, $u = \texttt{true}$), 
as they induce observable effect, %
and require access to the reference |y|. However, the function |usearg| can be reasoned in the mention mode by using \lemref{envt_strengthenW1} (shown below).

\bfparagraph{Properties} The following lemma allows reasoning about a sub-term in the mention mode, if the overall reasoning is in a use mode.
\begin{lemma}\label{envt_strengthenW1}
    If $(\W, \VE, \R) \in \VG{\G \ \flt \ u}$
    and $u'$ implies $u$,
    then $(\W, \VE, \R) \in \VG{\G \ \flt \ u'}$.
\end{lemma}

The following lemma allows the established equivalent terms in the mention mode to remain equivalent in the use mode,
provided they induce no observable effects and carry no abilities.
\begin{lemma}\label{exp_mentionable1}
    If $(\W, \VE, t_1, t_2, \texttt{false}) \in \ETAE{T}{a}{e}$, and 
    $e = \bfalseE$, and 
    $a = \avbot \vee u = \texttt{false}$,
    then $(\W, \VE, t_1, t_2, u) \in \ETAE{T}{a}{e}$.
\end{lemma}

\subsection{Value Interpretation of Types}\label{sec:valt}
\begin{figure*}[t] \small
    \begin{mdframed}%
 \judgement{Binary Value Interpretation of Types and Terms}{$\BOX{\lambdaae}$} \vspace{-2ex} 
 \begin{mathpar}
    \begin{array}{@{\hspace{-2ex}}r@{\hspace{1ex}}l@{\hspace{1ex}} l}
        \stw{\sigma}{\sigma'}{L}  & = & \forall \ell \in \DOM(\sigma). \ \ell \not\in L \Rightarrow \sigma(\ell) = \sigma'(\ell)  \\
        \fr{\sigma'}{\sigma} & = & \DOM(\sigma') - \DOM(\sigma) \\
        \\
        \VT{\TBoolAE}  & = & \{(\W, b, b, u, \lsv{1}, \lsv{2}) \} \\ 
        \VT{\TRefAE} & = & \{(\W, \ell_1, \ell_2, u, \lsv{1}, \lsv{2}) \mid  u \Rightarrow (\W(\ell_1, \ell_2) \ \land \ \ell_1 \in \lsv{1} \, \land \, \ell_2 \in \lsv{2}) \}  \\   
        \\
        \VT{\TFunAE{T_1}{a_1}{T_2}{a_2}{e}} & = & \{(\W, \cll{H_1}{\tabs{t_1}}, \cll{H_2}{\tabs{t_2}}, u, \lsv{\lambda_1}, \lsv{\lambda_2}) \mid  \forall \sigma_1', \sigma_2', \W', \lsv{1}, \lsv{2}, v_1, v_2, \lsv{v_1}, \lsv{v_2}. \\[\medskipamount]
                            & & \;\; \forall\, u_x, u_b, u_r. \, \NUM{L1}{((\ef{e} \vee u_r \wedge \ab{a_2.s}) \Rightarrow  \stcp{\W}{\W'}{\lsv{\lambda_1}}{\lsv{\lambda_2}})} \, \land \, \\[\medskipamount]
                            & & \qquad ((\ef{e}  \vee u_r \wedge \ab{a_2.s}) \Rightarrow (u_x \wedge u)) \, \land  \, (\ef{e} \Rightarrow u_b) \, \land \, (u_r = \neg a_1.s \vee u_b) \, \land \\[\medskipamount]
                            & & \qquad  (\ab{\neg \bar{a}_1} \Rightarrow u_x ) \, \land \, \DOM(\W) \subseteq \DOM(\W') \, \land \, \NUM{S1}{\st{\sigma_1'}{\sigma_2'}{\W'}{\lsv{1}}{\lsv{2}}} \, \land \, \\[\medskipamount]  
                            & & \qquad \NUM{R1}{(e = \Rightarrow \lsv{v_i} \cup \lsv{\lambda_i} \subseteq \lsv{i})} \, \land  \, \NUM{R2}{((\ab{\neg \bar{a}_1} \vee \neg u_x) \Rightarrow (\lsv{v_i} = \emptyset))} \, \land \\[\medskipamount]
                            & & \qquad (\W', v_1, v_2, u_x, \lsv{v_1}, \lsv{v_2}) \in \VT{T_1} \, \Rightarrow \\[\medskipamount]
                            & & \qquad (\exists \sigma_1'', \sigma_2'', \W'', r_1, r_2, \lsv{r_1}, \lsv{r_2}. \, \W' \sqsubseteq \W'' \, \land \, 
                                 \config{H_1\extends(x, v_1)}{\sigma_1'}{t_1} \eval \pconfig{\sigma_1''}{r_1} \, \land \\[\medskipamount]
                            & & \qquad \qquad \config{H_2\extends(x, v_2)}{\sigma_2'}{t_2} \eval \pconfig{\sigma_2''}{r_2} \, \land \, \DOM(\sigma_i') \subseteq \DOM(\sigma_i'') \, \land \, \\[\medskipamount]
                            & & \qquad \qquad \NUM{S2}{\st{\sigma_1''}{\sigma_2''}{\W''}{\lsv{1} \cup \fr{\sigma_1''}{\sigma_1'}}{\lsv{2} \cup \fr{\sigma_2''}{\sigma_2'}}} \, \land \, (\neg u_r \Rightarrow L_{r_i} = \emptyset) \, \land \\[\medskipamount]
                            & & \qquad \qquad (\W'', r_1, r_2, u_r, \lsv{r_1}, \lsv{r_2}) \in \VT{T_2} \, \land \, \\[\medskipamount]
                            & & \qquad \qquad \NUM{R3}{\lsv{r_i} \subseteq \emptyset \ccup{\ab{a_2.s}} (\lsv{v_i} \cup \lsv{\lambda_i}) \ccup{\ab{a_2.f}} \fr{\sigma_i''}{\sigma_i'}} \, \land \, \\[\medskipamount]
                            & & \qquad \qquad \NUM{E1}{\stw{\sigma_i'}{\sigma_i''}{\emptyset \ccup{\textcolor{red}{e}} (\lsv{\lambda_i} \cup \lsv{v_i})}} \, \land \, 
                                           \NUM{F1}{(\neg \ab{a_2.f} \Rightarrow \W''_f = \W'_f)})    
                            \}  \\
        \\
        \ETAE{T}{a}{e} & = & \{(\W, \VE, t_1, t_2, u) \| \forall \sigma_1, \sigma_2, \lsv{1}, \lsv{2}, \, \st{\sigma_1}{\sigma_2}{\W}{\lsv{1}}{\lsv{2}} \, \land \, \emptyset \ccup{\ef{e}} \explocs{t_i}{\VE_i} \subseteq \lsv{i} \, \land \, \\[\medskipamount]
                           &   & \;\; \exists \sigma_1', \sigma_2', \W', u', v_1, v_2, \lsv{v_1}, \lsv{v_2}. \, \config{\VE_1}{\sigma_1}{t_1} \eval \pconfig{\sigma_1'}{v_1} \,\land \,
                                 \config{\VE_2}{\sigma_2}{t_2} \eval \pconfig{\sigma_2'}{v_2} \, \land \, \\[\medskipamount]
                           &   & \;\; \qquad  \W \sqsubseteq \W' \, \land \,\DOM(\sigma_i') \subseteq \DOM(\sigma_i)\, \land \, \st{\sigma_1'}{\sigma_2'}{\W'}{\lsv{1} \cup \fr{\sigma_1'}{\sigma_1}}{\lsv{2} \cup \fr{\sigma_2'}{\sigma_2}} \, \land \, \\[\medskipamount]
                           &   & \;\; \qquad (\W', v_1, v_2, u', \lsv{1}, \lsv{2}) \in \VT{T} \, \land \, u' = (\neg \ab{\bar{a}} \vee u) \, \land \, (\neg u \Rightarrow L_{v_i} = \emptyset) \, \land \\[\medskipamount]
                           &   & \;\; \qquad \vallocs{v_i} \subseteq \emptyset \ccup{\ab{a.s}} \explocs{t_i}{\VE_i} \ccup{\ab{a.f}} \fr{\sigma_i'}{\sigma_i} \, \land \, \\[\medskipamount]
                           &   & \;\; \qquad \NUM{E2}{\stw{\sigma_i}{\sigma_i'}{\emptyset \ccup{\textcolor{red}{e}} \explocs{t_i}{\VE_i}}} \, \land \, \NUM{F2}{(\neg \ab{a.f} \Rightarrow \W'_f = \W_f)}
                           \} 
    \end{array}
\end{mathpar}       
\end{mdframed}
\caption{The binary interpretation of types and terms.}
\label{fig:binary}
\end{figure*} %
\figref{fig:binary} shows selected value interpretation of types. See the supplement for the complete definition.
In the figure, we use the subscript $i$ in formulas to refer to both terms when the context is clear,
where $i = 1$ or $2$. For example, %
we write $\lsv{v_i} = \emptyset$ to mean $\lsv{v_1} = \emptyset \, \land \, \lsv{v_2} = \emptyset$.

To simplify notation and improve readability, we identify the values of $\btrueA$/$\btrueE$ and $\bfalseA$/$\bfalseE$ with 
logical truth and falsehood, respectively.
For example, we write $e \Rightarrow P$ (for some effect $e$ and predicate $P$) to mean $e = \btrueE \Rightarrow P$.
In addition, we use the shorthand $\neg a$ to express $\neg (a.f \vee a.s)$,
meaning that a value does not reach any resource.
In \figref{fig:binary}, we use the notation $\ccup{b}$ to denote the conditional set union operator, \ie, if the predicate $b$ is true, 
the operator performs a union of the left- and right-hand operands;
otherwise, the result defaults to the left-hand operand only.

The value interpretation of types, written as $\VT{T}$, 
is a set of tuples of form $(\W, v_1, v_2, u, \lsv{1}, \lsv{2})$,
where $v_1$ and $v_2$ are values, and $\W$ is a world, 
$u$ denotes whether the pair of values can be used,
and $\lsv{1}$ and $\lsv{2}$ are some upper bound of the locations that the reasoning cares.

\bfparagraph{Boolean Type.} A pair of boolean values are related if they are both $\ttrue$ or $\tfalse$.
Since operations on boolean values do not need any resources, they can be used if the associated $u$ component allows.

\bfparagraph{Reference Type.} A pair of locations $(\ell_1, \ell_2)$ are related \wrt the world $\W$ if they are used by the reasoning (\ie, $u$ holds).
In this case, their relation must be defined in the world $\W$, and their reachable locations have to be bounded by $\lsv{1}$ and $\lsv{2}$ respectively.
Otherwise (\ie, $u$ does not hold), the reasoning does not have any knowledge about the pair.

\bfparagraph{Function Type.} 
The set of values of function type $\TFunAE{T_1}{a_1}{T_2}{a_2}{e}$ \wrt the world $\W$ is defined based on its computational behavior.
Note that we inline the term interpretation to characterize the function body's computational behavior.

\bfparagraph{Safe Execution.} 
As mentioned earlier, we adopt relational worlds from \citet{bao2025logrel}'s work to characterize localization.
For example, 
if a function either has a latent effect, 
or the result carries stored ability (\ie, $\ef{e} \vee u_r \wedge \ab{a_2.s}$),
then a safe execution of the function body needs to access locations reachable from the function (formulas $\recdx{L1}$ and $\recdx{S1}$).
Here, the formula $\stcp{\W}{\W'}{L_{\lambda_1}}{L_{\lambda_2}}$ means that types are only preserved \wrt 
reachable locations from function values.

Thus, the reasoning must need the locations reachable from the function.
In another word, if the function has no latent effect, 
or its return value does not reach any pre-existing locations (\ie, $\textcolor{blue}{a_2.s=\bfalseA}$), 
then the function can be safely executed in any future world extending from the current one $\W$ where the argument values are valid.

\bfparagraph{From Abilities and Effects to Reachable Locations.} 
Formulas $\recdx{R1}$, $\recdx{R2}$ and $\recdx{R3}$ capture properties of reachable locations enabled by 
the ability and effect qualifiers.
Formula $\recdx{R1}$ states that 
if the function has a latent effect, 
then the reachable locations from the function and its argument are bounded by the locations that the reasoning cares (\ie, $\lsv{i}$).
Formula $\recdx{R2}$ states that the argument cannot reach any locations 
if the fresh and stored component of the argument's ability qualifier are both $\bfalseA$ or the argument is not used.
Formula $\recdx{R3}$ states that if the return value's stored component (in its ability qualifier) is $\btrueA$, 
then it may reach locations from the function or its argument.
In addition, if its fresh component is $\btrueA$, it may reach freshly allocated locations.

\bfparagraph{From Effects to Value Preservation.}
Formulas $\recdx{E1}$ and $\recdx{E2}$ capture the value preservation over a potentially effectful computation,
\ie, values stored in pre-existing locations that are not covered by the effects remain unchanged. 

\bfparagraph{Terms.}
In addition to the properties about reachable locations, safe execution and value preservation,
which are similar to those for function bodies, 
two terms are related if their reduction results are related at given types.

\subsection{The Compatibility Lemmas, Fundamental Theorem and Soundness}

The semantic typing rules have the same shape as the corresponding syntactic ones, 
but turning the symbol $\vdash$ in \figref{fig:has-type-ae} into $\models$. 
Each of the semantic typing rules is a compatibility lemma. 

\begin{theorem}[Fundamental Property] If $\hasTypeAEC{\G} {t} {T} {a} {e}$, then $\semTypeAE{\G}{t}{t}{T}{a}{e}$.
\end{theorem}

\begin{theorem}[Soundness \wrt Observational Equivalence] The binary logical relation is sound \wrt observational equivalence, \ie,
    if $\hasTypeAEC{\G\ \ab{a_{\flt}}}{t_1}{T}{a}{e}$, and $\hasTypeAEC{\G\ \ab{a_{\flt}}}{t_2}{T}{a}{e}$, 
    then $\semTypeAE{\G}{t_1}{t_2}{T}{a}{e}$ implies $\ctxEquivAE{\G\ \ab{a_{\flt}}}{t_1}{t_2}{T}{a}{e}$.\footnote{For technical reason, our proof requires that $t_1$ and $t_2$ have the same set of free variables.}
\end{theorem}

\begin{theorem}[Effect Safety of Ability and Effect Type Systems]
    A syntactically pure expression is semantically (observationally) pure:
    $$ 
    \small
      \hasTypeAEC{\G\ \ab{a_{\flt}}}{t}{T}{a}{\bfalseE} \Rightarrow 
        \forall C: (\G\ \ab{a_{\flt}}; T\ \ab{a}\ \bfalseE) \carrow (\G'\ \flt'; T'\ \ab{a'}\ \ef{e'}).\
        \G' \models (\lambda x. C[x])\ t \equiva C[t]: T'\ \ab{a'}\ \ef{e'},
    $$
    where $\ab{a.f = \bfalseA}$, and $\ab{a <: a_{\flt}}$.%
\end{theorem}

\begin{theorem}[$\beta$-Equivalence]\label{thm:beta}
    If $\hasTypeAEC{\G, x: T_1\ a_1}{t_2}{T_2}{a}{e}$, 
    and $\hasTypeAEC{\G}{t_1}{T_1}{a_1}{\bfalseE}$,
    and $\ab{a_1.f = \bfalseA}$,
    then $\semTypeAE{\G}{(\tabs t_2)(t_1)}{t_2[t_1/x]}{T_2}{a}{e}$.
\end{theorem}

\begin{theorem}[Reordering]\label{thm:reorder}
    If $\hasTypeAEC{\G}{t_1}{\TBoolAE}{a_1}{e_1}$, and 
       $\hasTypeAEC{\G}{t_2}{\TBoolAE}{a_2}{e_2}$, and 
       $\ef{e_1 \wedge e_2 = \bfalseE}$, 
    then $\semTypeAE{\G}{\tseqb{t_1}{t_2}}{\tseqb{t_2}{t_1}}{\TBoolAE}{\avbot}{e_1 \EFC e_2}$
\end{theorem}
\section{Discussion and Possible Extensions}\label{sec:diss}%

\bfparagraph{Deterministic Evaluation}
In this paper, we assume that evaluation itself is deterministic, 
\ie, the evaluation order is fixed (\eg, left-to-right, call-by-value,  or any other deterministic convention). 
Allocations always pick the \emph{next} free store location rather than
any unused one ($\ell = \|dom(\sigma)\|$ in \figref{fig:syntax}). 
This simplifies the technical development without undue restriction, as many 
instances of nondeterminism can be made deterministic in a similar way.
For example, concurrency could be
modeled by an oracle that picks the next thread to execute
\cite{DBLP:conf/csl/BirkedalST12}.

\bfparagraph{Finer-Grained Store Effects} 
This work intentionally adopts a binary pure/impure effect system, 
as it suffices for our goal of studying the expressiveness of \emph{simple}
effect typing systems.
Extending the framework with a finer-grained effect system is straightforward. 
For example, one can refine the effect component with read and write, 
and the ability system can also be refined to read and write capabilities. 
Such refinements, however, introduce additional semantic complexity, 
\eg, modeling fine-grained effects typically requires explicit store invariants. 
By contrast, purity -- the absence of all effects -- can be captured purely observationally using our definition.

\bfparagraph{Multiple Effect Domains}
This work focuses on store effects as a representative effect domain, 
as mutable state has certain interesting properties, such as effect masking. 
Extending the system to support multiple, orthogonal effects (\eg, exceptions and I/O) 
would primarily require extending the language (with or without separate effect annotations), 
their associated typing rules (as mentioned in~\citet{DBLP:conf/ecoop/RytzOH12}), 
as well as proper characterization in the definition of logical relations, 
without fundamentally altering the structure of the system.

\bfparagraph{Taming Complexity}
There is a significant amount of complexity behind our formal results,
especially in the definition of the logical relation (\ie \figref{fig:binary})
and the semantic typing proofs.
We find it encouraging that despite this internal complexity, 
the complexity of $\lambdaae$ at the surface typing level (\ie \figref{fig:has-type-ae}) remains quite manageable for what the system achieves (\eg, effect polymorphism without explicit effect parameters). Most importantly, the types themselves remain relatively simple, with just boolean freshness, ability, and effect qualifiers.
Thus, we like to think the system achieves a good ``power-to-weight ratio''.
Despite this, efforts to simplify both internal invariants as well as user-facing typing rules would be promising research directions. %
\section{Related Work} \label{sec:related}

\bfparagraph{Purity} 
In object-oriented programming languages, 
various definitions of purity have been proposed~\cite{stewart2014csse}. 
\citet{DBLP:journals/sigsoft/LeavensBR06}'s definition has been widely adopted in Java:
a location is (observably) modified when it exists in both the pre-state and post-state 
of an execution, but stores different values in these two states.
This notion of purity as the absence of modification has been adopted by many Java tools, \eg, JPPA~\cite{DBLP:conf/vmcai/SalcianuR05}, ReIm and ReImInfer~\cite{DBLP:conf/oopsla/HuangMDE12,DBLP:conf/sigsoft/HuangM12},
JPure~\cite{DBLP:conf/cc/Pearce11}, as well as by Scala purity checking, \eg, \cite{DBLP:conf/ecoop/RytzAO13}.
By contrast, Joe-E~\cite{DBLP:conf/ndss/MettlerWC10} defines pure methods as being side-effect-free and deterministic, 
\ie, their outputs should depend only on their inputs. %
This notion of purity is closer to that in functional programming, \eg, it does not cover termination.

In this paper, we use the definition of a pure function aligning with the mathematical definition of a function. 
That is, a pure function's behavior is entirely determined by its input arguments, 
and its output does not depend on an external runtime environment that may change.
This definition is the most suitable for our purpose of equational reasoning. 

\citet{DBLP:journals/jfp/Sabry98} proposes to define purity of functional languages based on parameter-passing independence, 
\ie, different evaluation strategies (call-by-value, call-by-need, or call-by-name)
should lead to equivalent values modulo divergence and runtime errors.
In the age of Rocq, Agda~\cite{DBLP:conf/tphol/BoveDN09}, Lean~\cite{DBLP:conf/cade/Moura021}, and especially also Liquid Haskell~\cite{DBLP:phd/basesearch/Vazou16}, 
we take the position that it is useful to tighten the definition and
recognize nontermination as impurity in closer correspondence with
logical consistency. 
In the presence of possibly-infinite codata, 
a syntactic guardedness checker may be used to ensure productivity, \ie, even if a program generates an infinite amount of data, 
each piece will be generated in finite time~\cite{DBLP:journals/jucs/ATurner04}.

In the study of categorical semantics, an expression in an effectful language is pure if 
it can be substituted like a value. \citet{DBLP:conf/fossacs/Fuhrmann02}'s characterization
of such expressions as \emph{algebraic values} coincides with our notion of observational purity.
The similar notion of thunkable expression/morphism~\cite{Hayothesis,DBLP:journals/entcs/Fuhrmann99} is perhaps more suitable for a language that implements effects as a monad, or a categorical model of the semantics, whereas our notion of purity (\defref{def:opurety}) is defined purely in terms of observational equivalence with respect to an operational semantics.

\bfparagraph{Expressiveness}
\citet{DBLP:journals/scp/Felleisen91} formalizes a theory of expressiveness for 
programming languages based on structure-preserving translations
and distinguishes \emph{macro expressibility}, 
which requires the translation of each eliminated construct to be compositional. 
When considering type systems specifically, a number of finer-grained notions
of expressiveness have been proposed, including \citet{DBLP:books/sp/LongleyN15}'s
higher-order computability. \citet{DBLP:journals/lmcs/Longley18} shows that
while adding any fixpoint combinator to PCF makes the language Turing-complete,
higher-order fixpoint combinators can type strictly more terms. Precisely,
a fixpoint operator for a type of level $k$ cannot be defined using fixpoint 
operators for lower types. \citet{DBLP:journals/jfp/HillerstromLL24} bring
complexity into consideration as a measure of expressiveness, and 
(similar to our work) study expressiveness of effectful features.
Specifically, they deliver a precise account of how multi-shot continuations 
are more expressive than single-shot continuations, because multi-shot continuations 
can express asymptotically more efficient search strategies than single-shot continuations.

\bfparagraph{Type-and-Effect Systems} 
Effect systems were introduced to integrate imperative operations into functional languages~\cite{gifford1986integrating,DBLP:conf/popl/LucassenG88},
to track read and write effects on the heap~\cite{nielson1999type}, 
to check purity~\cite{DBLP:conf/cc/Pearce11,DBLP:conf/ecoop/RytzAO13} and exceptions~\cite{DBLP:conf/popl/PessauxL99,DBLP:journals/toplas/LeroyP00,java}, control effects~\cite{DBLP:conf/aplas/AsaiK07,DBLP:conf/icfp/RompfMO09},
and others. 
\citet{DBLP:conf/tldi/MarinoM09} introduced a generic type-and-effect system that allows programmers to easily define new effects systems. 
\citet{DBLP:conf/popl/Tate13} proposed a general framework for modeling sequential, \ie, flow-sensitive effects.
\citet{DBLP:journals/toplas/Gordon21} introduced the effect quantale framework
that can model the effects of a wide range of sequential effect systems. 
In this work, we adopt a canonical binary effect system that distinguishes between computations that may induce observable effects and those that do not,
serving the purpose of reasoning about purity as the absence of \emph{any} observable effect.

\bfparagraph{Algebraic Effects and Handlers} 
Algebraic effects~\cite{DBLP:journals/acs/PlotkinP03} and effect handlers~\cite{DBLP:journals/corr/PlotkinP13} are purely functional, composable approaches to modeling effects,
and inherit similar challenges in supporting effect polymorphism as those encountered in traditional effect systems.
These challenges are evident in languages that support algebraic effects, 
\eg, Koka~\cite{DBLP:conf/popl/Leijen17}, Helium~\cite{DBLP:journals/pacmpl/BiernackiPPS20}, the language in \citet{DBLP:conf/rta/HillerstromLAS17}'s work,
and the system developed by \citet{DBLP:journals/pacmpl/ZhangM19}'s work.
Frank~\cite{DBLP:conf/popl/LindleyMM17,DBLP:journals/jfp/ConventLMM20} reduces the annotation burden by leveraging ambient effects provided by the context, 
but relies on desugaring them to traditional effect polymorphism. Still, this mechanism has been described as fragile, potentially resulting in unclear error messages~\cite{DBLP:journals/pacmpl/TangWDHLL25}.
\citet{DBLP:journals/pacmpl/TangWDHLL25,tang2025rowscapabilitiesmodaleffects} extend Frank's approach with multimodal type theory~\cite{DBLP:journals/lmcs/GratzerKNB21,694f77d247d34986bb82129b8d96206e} 
in tracking modes for types and terms.
Similar to our work, \citet{tang2025rowscapabilitiesmodaleffects} compare row-based and capability-based effect systems 
by translating them (via macros) into a uniform framework $\mathrm{Met}(\mathcal{X})$.
Their language can express the richer notion of modal effects,
and treats user-defined effects using effect handlers. 
Our $\lambdaae$ combines abilities and built-in store effects in order to study purity.
They prove syntactic effect safety by syntactic means, whereas 
we prove effect safety from the perspective of 
semantic purity via logical relations.

\bfparagraph{Capabilities} 
Capability systems~\cite{DBLP:conf/ecoop/CastegrenW16,DBLP:conf/ecoop/BoylandNR01} have been used to reason about program resources
and external calls~\cite{drossopoulou2025reasoning},
and to ensure unique access with borrowing~\cite{haller2010capabilities}.
Reference capabilities in Pony~\cite{DBLP:phd/ethos/Clebsch17} distinguish read and write capabilities.
\citet{DBLP:conf/ecoop/Gordon19} articulates the use-mention distinction when designing capability and effect systems. 
Our $\lambdaae$ system uses ambient ability to track ``mention'', and effects to track ``use'', where effects are bounded by the ambient ability. 
\citet{DBLP:conf/icfem/CraigPGA18} use capabilities to bound the effects that an expression can induce for a language with 
a fixed set of global resources. Our systems can reason about dynamically allocated resources, too. 
\citet{osvald2016gentrification} distinguish 1st- and 2nd-class values via privilege qualifiers,
which can mediate access to different kinds of capabilities (\eg, read and write) with a fine-grained privilege lattice.
Effekt~\cite{DBLP:journals/pacmpl/BrachthauserSO20} treats effects as capabilities and treats all functions as second-class values,
which provides a lightweight form of effect polymorphism, though impeding reasoning about purity.
Our systems track resources via abilities in a coarse-grained way and can recognize pure computation, 
which was not a goal of those prior systems.

\bfparagraph{Coeffects} Coeffects~\cite{DBLP:conf/icalp/PetricekOM13} 
track what computational resources a computation requires from its context, similar 
to the ambient ability in our systems $\lambdaa$ and $\lambdaae$. However,
coeffects have no inherent notion corresponding to the capturing of ability from the context, 
which is essential, \eg, for our support for effect masking.
\citet{DBLP:conf/esop/BrunelGMZ14,granule,DBLP:conf/icfp/PetricekOM14} combine bounded linear coeffects with grades, enabling reasoning about the number of times a variable is used during evaluation. 
These works do not distinguish using a capability to cause an effect from just mentioning it~\cite{DBLP:conf/ecoop/Gordon19}.
In contrast, our $\lambdaae$ does not enforce linearity but distinguishes mention from use. 
\citet{DBLP:conf/icfp/GaboardiKOBU16} generalize a linear coeffect calculus to a combined effect-coeffect system, 
in which pure terms correspond to the standard terms of the $\lambda$-calculus.

\bfparagraph{Call-by-Push-Value}\citet{DBLP:phd/ethos/Levy01} introduced call-by-push-value (CBPV), 
which makes a distinction between value types (strict, effect-free) and computation types (lazy, side-effecting).
\citet{DBLP:journals/pacmpl/TorczonAAVW24} adapt and extend CBPV to track a single effect (tick) and single coeffect (resource usage),
where effects are modeled by a preorderd monoid~\cite{DBLP:conf/popl/Katsumata14}.
\citet{DBLP:conf/popl/Levy17} studies type isomorphisms using contextual morphism,
and uses thunkable maps for value types and linear maps~\cite{DBLP:conf/fossacs/Munch-Maccagnoni14} for computation types.
We study purity using operational semantics, and define contextual equivalence of typed terms.

\bfparagraph{Tracking Variables in Types} 
\citet{DBLP:journals/pacmpl/BaoWBJHR21} introduce reachability types to track aliasing and separation through reachability qualifiers.
Their work has been extended to support polymorphism~\cite{DBLP:journals/pacmpl/WeiBJBR24},
cyclic references~\cite{deng2025completecyclereachabilitytypes}, arenas~\cite{he2026arenas}, move semantics~\cite{DBLP:journals/corr/abs-2510-08939} and bidirectional typing~\cite{jia2026escape}.
Our semantic model is adapted from \citet{bao2025logrel}'s work, but simplified to Boolean references. 
\citet{bunting2025a} study contextual equivalence for \citet{DBLP:journals/pacmpl/BaoWBJHR21}'s system  
using game semantics and present a full-abstraction model based on a labelled transition system.

Capturing types~\cite{DBLP:journals/toplas/BoruchGruszeckiOLLB23,DBLP:journals/pacmpl/XuBO24} 
integrate capability tracking and escape checking into Scala 3.
Treating effects as capabilities provides a lightweight way to achieve effect polymorphism~\cite{DBLP:journals/pacmpl/BrachthauserSO20}.
\citet{DBLP:conf/programming/XuO24} introduce reach capabilities to address the challenges when capture checking and mutable variables interact.

We believe that the techniques presented here will be helpful in firmly characterizing what purity guarantees 
reachability and capturing type systems can make and for studying differences in expressiveness between such 
related systems.
Integrating mechanisms from our approach could potentially inform more precise purity checking for 
languages such as Scala.

\vspace{-1ex}
\section{Conclusions} \label{sec:conc}
\vspace{-1ex}

This paper proposed a semantic framework for purity and a measure of expressiveness for effect typing disciplines. We showed that canonical binary effect and capability systems are incomparable, motivating a synthesis that combines their strengths, which we call \emph{type, ability, and effect systems}. Our semantic definition of purity, along with our mechanized logical relation, provides a foundation for comparing and proving properties across diverse effect typing disciplines.
\bfparagraph{Data-Availability Statement}  
Rocq mechanizations can be found at \url{https://github.com/tiarkrompf/reachability} and~\cite{bao_2025_16934167}.
 
\begin{acks}                            %
  This paper has benefited from fruitful discussions 
  at the 24th Meeting of IFIP WG 2.11 in Edinburgh with Jeremy Gibbons, 
  Sam Lindley, and others.
  The paper has also profited from additional discussions with Ross Tate, Deivid Vale,
  and the anonymous reviewers.
  This work was supported in part by NSF award 2348334 and an Augusta faculty startup
  package, as well as funds from Purdue and gifts from Meta, Google, Microsoft, and VMware.
\end{acks}

\bibliography{references}

\end{document}